\documentclass[twocolumn,showpacs,preprintnumbers,amsmath,amssymb]{revtex4}
\usepackage{graphicx,subfigure}
\usepackage{color}
\pdfoutput=1
\newcommand {\bea} {\begin{eqnarray}}
\newcommand {\eea}{\end{eqnarray}}
\newcommand{\bem}{\begin{multline}}
\newcommand{\eem}{\end{multline}}
\newcommand {\beq} {\begin{equation}}
\newcommand{\ddx}{\hspace{-5.mm}dx}
\newcommand {\eeq}{\end{equation}}

\newcommand{\dP}{\int\hspace{-2.mm}\frac{d{\bf P}}{(2\pi)^3}}
\newcommand{\dk}{\int\frac{d{\bf k}}{(2\pi)^3}}
\newcommand{\dq}{\int\frac{d{\bf q}}{(2\pi)^3}}
\newcommand{\ds}{\int_{\mathcal C_{\gamma}\hspace{-2.mm}}\frac{ds}{2\pi i}}

\newcommand{\dppp}{\int\hspace{-1.mm}\frac{d{\bf P}_t d{\bf p'}_1}{(2\pi)^6}\int_0^{+\infty}\hspace{-5.mm}dx}
\newcommand{\dpp}{\int\hspace{-1.mm}\frac{d{\bf p'}_1 d{\bf p'}_2}{(2\pi)^6}\int_0^{+\infty}\hspace{-5.mm}dx}
\newcommand{\dpppp}{\int\hspace{-1.mm}\frac{d{\bf P}'d{\bf p'}_1}{(2\pi)^6}\int_0^{+\infty}\hspace{-5.mm}dx}
\newcommand{\dpppbis}{\int\hspace{-1.mm}\frac{d{\bf P}' d{\bf p'}_1}{(2\pi)^6}}
\begin{document}
\draft
\title{High-temperature expansion for interacting fermions}
\author{Mingyuan Sun$^{1,2}$ and Xavier Leyronas$^1$}
\affiliation{
$^1$ Laboratoire de Physique Statistique, Ecole Normale Sup\'erieure, UPMC
Univ Paris 06, Universit\'e Paris Diderot, CNRS, 24 rue Lhomond, 75005 Paris,
France\\
$^2$ Department of Physics and William Mong Institute of Nano Science and Technology, Hong Kong University of Science and Technology,
Clear Water Bay, Kowloon, Hong Kong}

\begin{abstract}
We present a general method for the high-temperature expansion of the self-energy of interacting particles. Though the method is valid for fermions and bosons, we illustrate it for spin one half fermions interacting  via a zero range potential, in the Bose Einstein Condensate - Bardeen Cooper Schrieffer (BEC-BCS) crossover.
The small parameter of the expansion is the fugacity $z$. Our results include terms of order $z$ and $z^2$, which take into account respectively two and three body correlations. 
We give results for the high temperature expansion of Tan's contact at order $z^3$ in the whole BEC-BCS crossover.
We apply our method to calculate the spectral function at the unitary limit. We find new structures which were overlooked by previous approaches, which included only two body correlations. 
This shows that including three-body correlations can play an important role in the structures of the spectral function.

\end{abstract}
\pacs{03.75.Hh, 03.75.Ss, 67.85.Pq}

\maketitle

\section{Introduction}
It is a general trend of the physics of ultra cold atoms to become the field of experiments simulating condensed matter problems \cite{rmpbdz}. In particular, interacting ultra cold fermionic atoms can be used to simulate interacting electrons in solid state physics. What is remarkable is that there is essentially no unknown relevant microscopic parameters and that some ({\it e.g.} the strength of the interaction) can be changed at will experimentally. Therefore the comparison to theory can be very accurate. An example is the measurement of the equation of state for ultra cold $^6Li$ atoms \cite{natureeoslkbli6,natphysmitfeynman}. In particular, the high temperature equation of state can be very well fitted using a virial expansion \cite{natureeoslkbli6}. At unitarity (where the scattering length diverges), the third order virial coefficient determined theoretically \cite{huliub3,b3blume,pravirial,praVNMMPJL} fits perfectly experimental results. 
More recently, the method of Ref.\cite{pravirial} was extended to the expansion of the self-energy in the context of the $2D$ BEC-BCS crossover \cite{parish-levinsen,barth-hofmann}. There the question of a depletion (the so-called pseudogap) in the spectral function, which physically represents the density of states of excitation energies after the creation or annihilation of a particle of wave vector $k$ was asked. Doing a lowest order calculation, which is a high temperature expansion of the $T$-matrix approximation, the authors of \cite{parish-levinsen,barth-hofmann} found that such a depletion can be found. However, the peculiarity of the $2D$ BEC-BCS crossover is that there exists always a two-body bound state called a dimer. Therefore it is quite natural (see Appendix \ref{appendixBEClimit} for a physical explanation of this fact) to find a pseudogap feature in the whole crossover. This is of course not the case in the $3D$ BEC-BCS crossover at the unitary limit, where there is no dimer state in the two body problem.  A high temperature expansion of the one-particle Green's function in a trap including $2$-body correlations was also studied in \cite{huliudos}. Using the $T$-matrix approximation, it was found at the unitary limit that there exists a depletion of the spectral function for moderate values of the wave vector $k$ \cite{pieristrinatipg,prlpieristrinatijin}.  Using a different theoretical approach, the spectral function was also studied in \cite{JK}. Experimentally, the observation of a pseudo gap has been found in Refs.\cite{djincamerino,prldjinnoFL}. This result was also found in some numerical calculations in Ref.\cite{prlbulgac} (which are numerically difficult since one has to find a real frequency quantity from imaginary time calculations). We recover this behavior in section \ref{Sigma1}, because our expansion at this order coincides with the high temperature behavior of the $T$-matrix approximation.
Our goal is to go beyond this approximation, in a high temperature controlled calculation. Indeed most of the analytical calculations of the spectral function to the best of our knowledge, have been done using this approximation.
So, we present in section \ref{secsigma2} our analytical expressions and in section \ref{numUL} our numerical results  beyond this approximation ({\it i.e} including three body correlations) at the unitary limit. 

The paper is organized as follows. In section \ref{general}, the general formalism is introduced. Then we present results of the high-temperature expansion for the self-energy, spectral function and Tan's contact to the lowest order in section \ref{Sigma1}. Section \ref{contact3} is devoted to a virial like calculation of Tan's contact at third order. In section \ref{secsigma2}, we give expressions for the fermion self-energy at next order in the fugacity, including $3$-body correlations. The technical details leading to the analytical expressions are explained in Appendices \ref{appsigdef} and \ref{eqT2T2p}.
In section \ref{numUL}, we show our numerical results for the spectral function  including three body correlations, pointing out the differences three body correlations make compared to the usual $T$-matrix approach (see Ref.\cite{pieristrinatipg} for instance).
Finally, we conclude in section \ref{conclusion}.

In this work, we will take the Boltzmann constant $k_B$ and the Planck's constant $\hbar$ equal to unity.
\section{General formalism}\label{general}
The general idea of the high temperature expansion \cite{pravirial} is to expand the fermionic non-interacting Green's functions in powers of the fugacity $z=e^{\beta\mu}$. Here, $\beta=1/T$ ($T$ the temperature) and $\mu$ is the chemical potential. If we denote by $G^{(0)}({\bf k},\tau)$ the non-interacting fermionic Green's function, we have
\begin{multline}
G^{(0)}({\bf k},\tau)=e^{-(\varepsilon_{{\bf k}}-\mu)\tau}
\left\{-\Theta(\tau)+n_F(\varepsilon_{{\bf k}}-\mu)
\right\}\label{eqG0in}
\end{multline}
with $\Theta(x)$ being the Heaviside function, $n_F(x)=1/(e^{\beta x}+1)$ is the Fermi-Dirac distribution and $\varepsilon_{{\bf k}}=k^2/(2m)$ ($m$ the mass of an atom). $\tau$ is the imaginary time. Next, the Fermi-Dirac distribution is expanded in powers of the fugacity : $n_F(\varepsilon_{{\bf k}}-\mu)=\sum_{n\geq 1}(-1)^{n+1}z^n e^{-n\beta \varepsilon_{{\bf k}}}$. We can write 
\bea
G^{(0)}({\bf k},\tau)=e^{\mu\tau}\left[
\sum_{n\geq 0} 
G^{(0,n)}({\bf k},\tau)\,z^n
\right]\label{eqG0}
\eea
 We have defined
\bea
G^{(0,0)}({\bf k},\tau)&=&-\Theta(\tau)e^{-\varepsilon_{{\bf k}}\tau}\label{eqg00}\\
G^{(0,n)}({\bf k},\tau)&=&(-1)^{n-1}e^{-\varepsilon_{{\bf k}}\tau}e^{-n\beta\varepsilon_{{\bf k}}},\,n\geq 1\label{eqg0nf}
\eea
Therefore $G^{(0,0)}$ is {\it retarded}, while $G^{(0,n)}$, for $n\geq 1$, is {\it not retarded}.
Notice that $G^{(0,n)}$ does not depend on the chemical potential $\mu$. Diagrammatically, since $G^{(0,0)}$ is a retarded function, we represent it as a line with an arrow going {\it from left to right} if increasing time goes to the right (this is the Green's function of a particle in vacuum).
On the other hand, $G^{(0,n)}$, for $n\geq 1$, is not retarded, and we represent it as a  $n$-times slashed line, which can be oriented from left to right or {\it vice versa}.  This formalism was used in \cite{pravirial} in order to calculate the fermion occupation number $n_k$, which is obtained by closing in time the exact Green's function. By integration on wave-vectors, we get the density $n(\mu,T)$ and by the Gibbs-Duhem relation, we can calculate the equation of state $P(\mu,T)$ where $P$ is the pressure of the system. This low fugacity expansion is a high-temperature low density expansion, as can be the low density expansion in a Lee-Huang-Yang like calculation for a molecular BEC \cite{prllhy,pralhy}. However, here the density is low compared to $\Lambda_T^{-3}$  where $\Lambda_T=\sqrt{2\pi/(m T)}$ is the thermal de Broglie wavelength. Instead in \cite{prllhy,pralhy}, the density is small compared to $a^{-3}$ where $a$ is the scattering length.

In this work, we use the high temperature expansion in order to calculate the retarded self-energy $\Sigma_R(k,\omega)$ in powers of the fugacity $z$. This allows us to calculate the retarded Green's function $G_R(k,\omega)$ and the spectral function $A(k,\omega)=-1/\pi \Im(G_R(k,\omega))$.

This approach was used in \cite{parish-levinsen} and \cite{barth-hofmann} at lowest order in the fugacity $z$ in the context of the $2D$ BEC-BCS crossover.
Our present calculation is instead up to order $z^2$ for the $3D$ BEC-BCS crossover. In particular, we take into account the $3$-body problem.
\section{Calculation of $\Sigma^{(1)}$}\label{Sigma1}

\begin{figure}[h]
\begin{center}
\includegraphics[width=.7\linewidth]{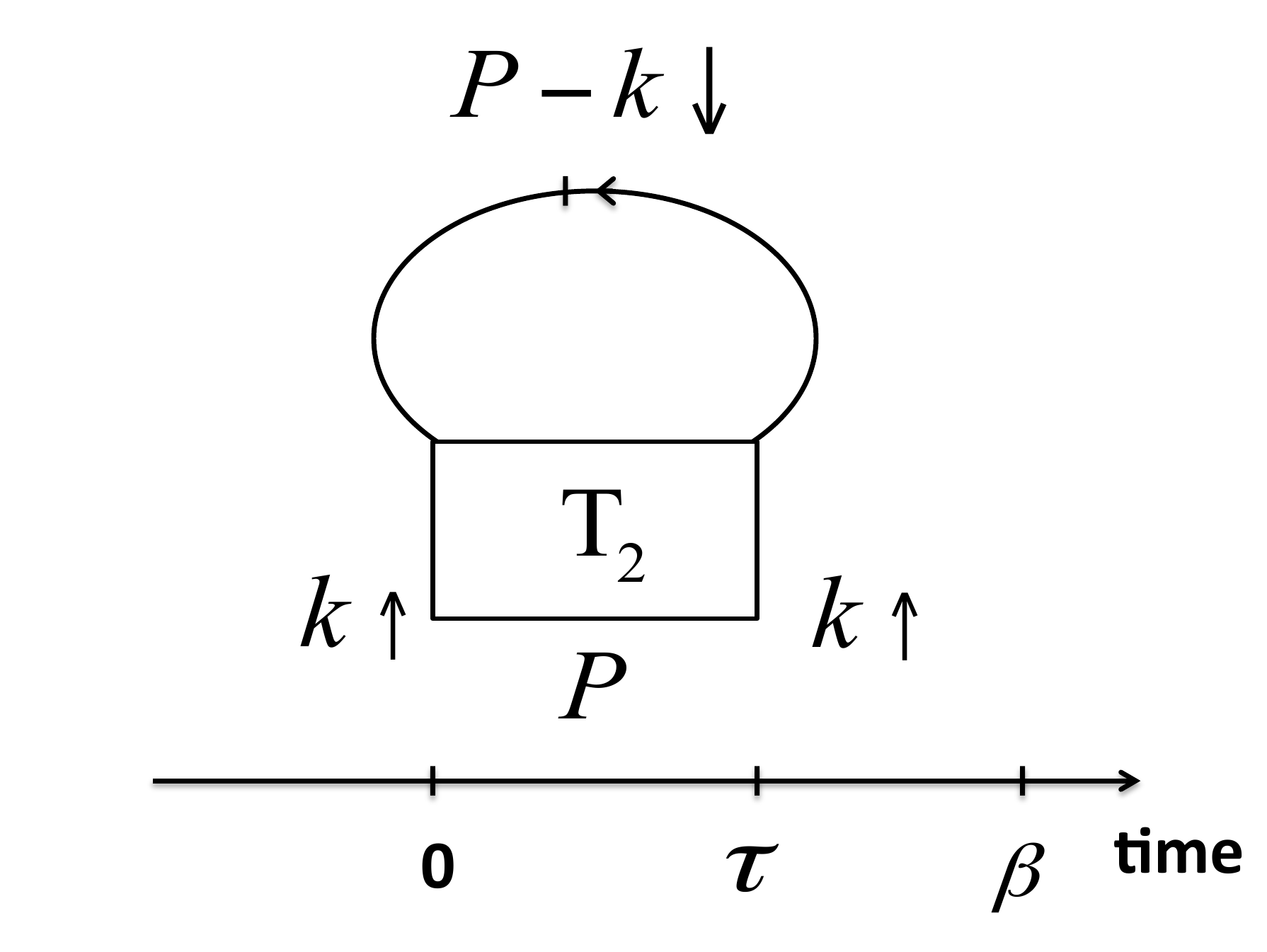}
\caption{Lowest order diagram for the self-energy.}
\label{figsigma1}
\end{center}
\end{figure}

The lowest order diagram for the self-energy is shown in Fig.\ref{figsigma1}. Previous authors \cite{parish-levinsen,barth-hofmann} have used this diagram in the context of the $2D$ BEC-BCS crossover. The analytical expression is
\bea
\Sigma^{(1)}(k,\tau)&=&z\dP e^{\mu\tau}e^{-(\beta-\tau)\varepsilon_{{\bf P-k}}}T_2(P,\tau)\label{eqsigma1_tau}
\eea
$T_2$ denotes the $2-$body $T-$matrix in the case of a contact interaction. In terms of Feynman diagrams, it is given by the sum of ladder diagrams \cite{pravirial}.
We use the Galilean invariance property $T_2(P,\tau)=e^{-\tau P^2/4m}T_2(0,\tau)$, and we denote $t_2(s)$ the Laplace transform of $T_2(0,\tau)$.
In $3D$ for equal masses, we have $t_2(s)=\frac{4\pi}{m}\left[a^{-1}-\sqrt{-m\,s}  \right]^{-1}$.
We find 
\bea
T_2(P,\tau)&=&e^{-\frac{P^2}{4 m}\tau}\ds e^{-\tau s}t_2(s)\label{eqT_2}
\eea where $\mathcal{C}_{\gamma}$ is a Bromwich contour \cite{appel2007mathematics}. $\gamma$ is such that the integrand is analytic for $s$ such that $\Re(s)<\gamma$. For $a^{-1}\leq 0$, it is sufficient to have $\gamma<0$, while for $a^{-1}>0$, $\gamma <-E_b$ due to the molecular pole $-E_b=-1/(m a^2)$ of $t_2$. In Eq.(\ref{eqT_2}), we deform the contour along the real axis and we find
\bea
T_2(P,\tau)&=&-e^{-\frac{P^2}{4 m}\tau}[
\Theta(a^{-1})Z_m e^{E_b \tau}\nonumber\\ 
&&+\int_0^{+\infty}\hspace{-3.mm}dx\,e^{-\tau x}\rho_2(x)
]\label{eqT_2_tau}
\eea
where $\Theta(x)$ is the Heaviside step function, $Z_m=8\pi/(m^2 a)$ is the molecular residue and  $\rho_2(x)=-1/\pi \Im(t_2(x+i 0^+))=4/m^{3/2}\sqrt{x}/(x+E_b)$ is the spectral density of the $2$-particle $T$-matrix.
The physical interpretation of the two terms of Eq.(\ref{eqT_2_tau}) is clear: the  first term comes from the two-body bound state of energy $-E_b$ (the dimer), while the second is due to the two-particle scattering states continuum of energy $x$. Using this expression in Eq.(\ref{eqsigma1_tau}), we can easily perform the Fourier transform 
$\int_0^{\beta}d\tau e^{i\omega_n\tau}\cdot$, with $\omega_n=\pi T(2 n+1)$ a fermionic Matsubara frequency.
We find
\beq
\Sigma^{(1)}(k,i\omega_n)=z\,F_1(k,i\omega_n+\mu)+z^2\,H_1(k,i\omega_n+\mu)
\eeq
where the two functions $F_1$ and $H_1$ are defined through
\begin{multline}
F_1(k,E)=\dP e^{-\beta\frac{({\bf P-k})^2}{2 m}}
[\int_0^{+\infty}\ddx
\frac{\rho_2(x)}
{
E-(\frac{P^2}{4m}+x-\frac{({\bf P-k})^2}{2 m})
}\\
+\Theta(a^{-1})\frac{Z_m}{E-(\frac{P^2}{4m}-E_b-\frac{({\bf P-k})^2}{2 m})}]\label{eqF_1}
\end{multline}
\bea
H_1(k,E)&=&\dP [\int_0^{+\infty} \ddx e^{-\beta(\frac{P^2}{4m}+x)}\frac{\rho_2(x)}{E-(\frac{P^2}{4m}+x-\frac{({\bf P-k})^2}{2 m})}\nonumber\\
&&+\Theta(a^{-1})\frac{Z_m e^{-\beta(\frac{P^2}{4m}-E_b)}}{E-(\frac{P^2}{4m}-E_b-\frac{({\bf P-k})^2}{2 m})}]\label{eqH_1}
\eea
for a general complex energy $E$ outside the real axis.
These expressions call for a simple physical interpretation. They contain energy denominators of the form $E-\Delta E$ where $\Delta E=P^2/(4m)+x-({\bf P-k})^2/(2m)$ is clearly the energy of an intermediate state with a spin up and a spin down interacting particles of center of mass kinetic energy $P^2/(4m)$, total momentum ${\bf P}$ and relative motion energy $x$, plus a spin down hole energy $-({\bf P-k})^2/(2m)$. This can be seen directly on the diagram of Fig.\ref{figsigma1}, where the two interacting particles are represented in the $T_2$ symbol which propagates forward in time, while the hole, represented by the slashed line propagates backward in time.  Terms of the type
$1/(E-(P^2/(4m)-E_b-({\bf P-k})^2/(2 m)))$ of course represent intermediate states where the relative motion is in the two-body bound state of energy $-E_b$.

From the expressions Eqs.(\ref{eqF_1}) and (\ref{eqH_1}), we can easily perform the analytical continuation to the real axis so as to get the retarded self-energy. Indeed, we see that $F_1$ and $H_1$ are analytic functions of $E$ in the upper and lower half complex plane. Therefore we find for the retarded self-energy
\bea
\Sigma^{(1)}_R(k,\omega)&=&z\, F_1(k,\omega+\mu+i 0^+)+z^2 \,H_1(k,\omega+\mu+i 0^+)\nonumber\\\label{eqsigma_1}
\eea 
From this, we come to the conclusion that the diagram of Fig.\ref{figsigma1} actually contains terms of order $z$ {\it and} $z^2$.
We can actually simplify the expression for $F_1$. We use the relation for the function $t_2(z)$
\bea
t_2(z)&=&\Theta(a^{-1})\frac{Z_m}{z+E_b}+\int_0^{+\infty}dx\frac{ \rho_2(x)}{(z-x)}
\eea
which comes from the analytic properties of $t_2(z)$ and which is therefore valid in any dimension (with the proper $Z_m$). Then we see from Eq.(\ref{eqF_1}) that
\bea
F_1(k,E)&=&\dP e^{-\beta\frac{({\bf P-k})^2}{2 m}} t_2(E-\frac{P^2}{4 m}+\frac{({\bf P-k})^2}{2 m})\nonumber\\
\eea
This is the result of Eq.($13$) of Ref.\cite{barth-hofmann} which was found for a $2D$ system. It was also derived in Ref.\cite{nishidaESR} in the context of the physics of magnons.
\subsection{Examples of $A(k,\omega)$}
\begin{figure}[h]
\subfigure[\label{spectralfunction+1}]{\includegraphics[width=0.75\linewidth]{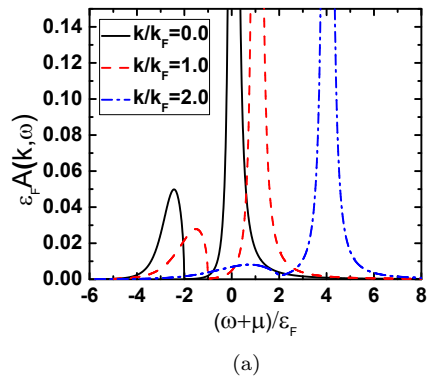}}\\
\subfigure[\label{spectralfunction0}]{\includegraphics[width=0.75\linewidth]{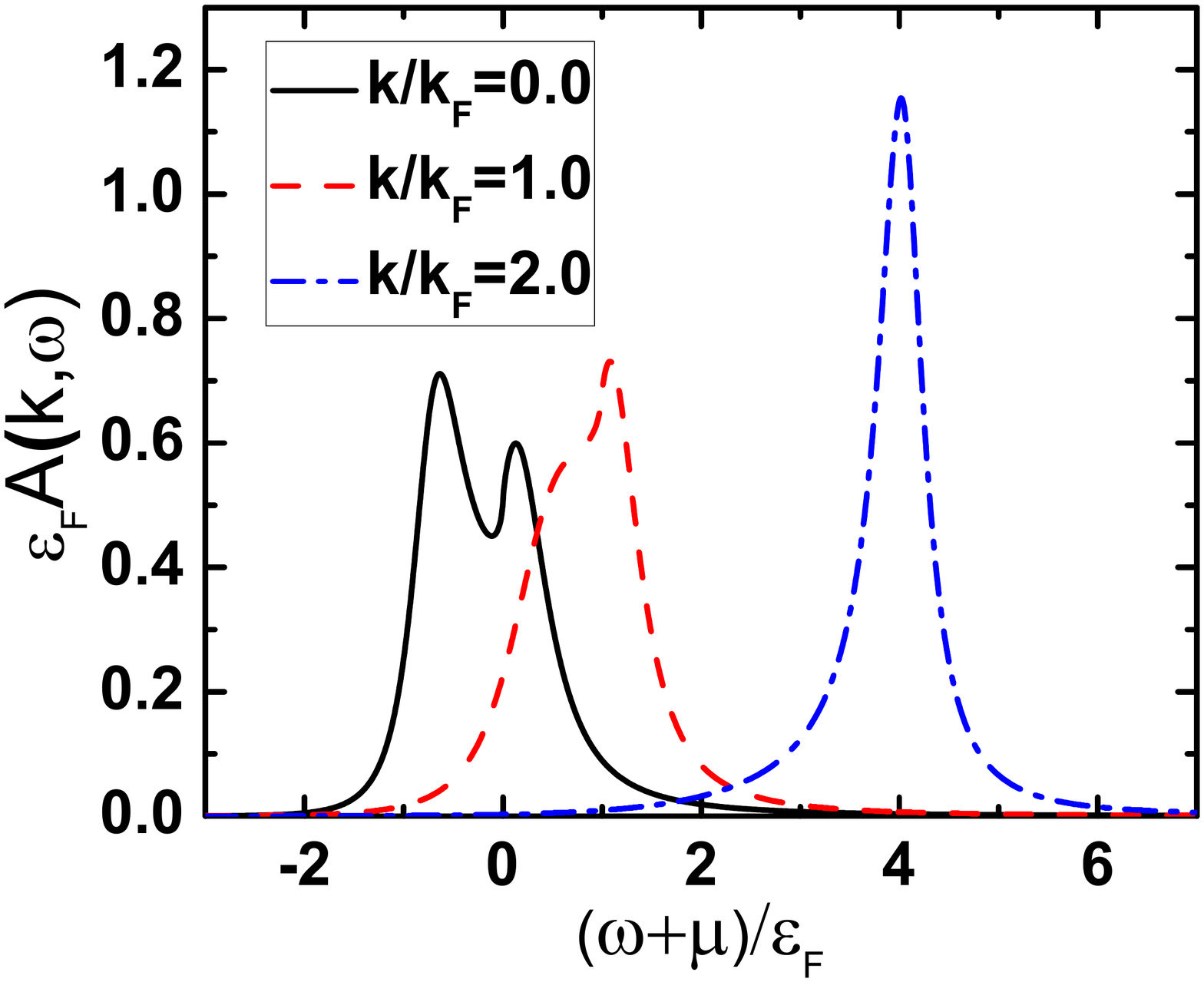}}\\
\subfigure[\label{spectralfunction-1}]{\includegraphics[width=0.75\linewidth]{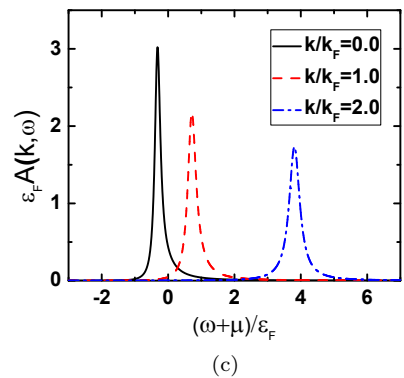}}
\caption{
Spectral functions $A(k,\omega)$ versus $\omega+\mu$ at $T=T_F$. (a) $1/(k_F a)=1$ (BEC limit), (b) $1/(k_F a)=0$ (Unitary Limit) and  (c) $1/(k_F a)=-1$ (BCS limit). In each panel, three values of the wave vectors are shown: $k/k_F=0$ (black solid line), $k/k_F=1$ (red dashed line) and $k/k_F=2$ (blue dash dotted line).}
\end{figure}
We compute the spectral function $A(k,\omega)$ from the identity
\bea
A(k,\omega)&=&-\frac{1}{\pi}\Im\left(G_{R}(k,\omega)\right)
\eea
The retarded Green's function is deduced from the computed retarded self-energy $\Sigma_R(k,\omega)$ via the Dyson equation
\bea
G_{R}(k,\omega)&=&\frac{1}{\omega+\mu+i0^{+}-k^2/(2m)-\Sigma_{R}(k,\omega)}
\eea
We show in Figs.\ref{spectralfunction+1},\ref{spectralfunction0} and \ref{spectralfunction-1} typical examples of spectral functions $A(k,\omega)$ using the lowest order self energy of order $z$ in Eq.(\ref{eqsigma_1}). This is the $3D$ version of what was done in $2D$ in \cite{parish-levinsen,barth-hofmann}.  However, contrary to the $2D$ case, in $3D$ there is no molecular state for $a^{-1}\leq 0$. We define the Fermi momentum $k_F$ in terms of the total particle density $n=k_{F}^3/(3\pi^2)$.
The Fermi energy $\varepsilon_F=k_F^2/(2m)\equiv T_F$. It is explained in Appendix \ref{appfug} how to determine the fugacity $z$ for given $T/T_F$ and $1/(k_F\,a)$.

Starting from the BEC side ($1/(k_F a)=1$, $z=0.17$), we find that the spectral function has in general two distinct features (see Fig.\ref{spectralfunction+1}: a peak on the positive $\omega+\mu$ part, centered around $\omega+\mu=k^2/(2m)$ (essentially a Lorentzian shape) and a  broad structure starting below $-E_b+k^2/(2m)$. These structures have a very simple physical interpretation. The peak centered around $k^2/(2m)$ corresponds to the creation of a spin $\uparrow$ fermion of momentum ${\bf k}$ in a gas of classical dimers. In the BEC limit $a$ is small and the scattering of the created fermion on the dimers is weak. These enable to find the position of the peak, with a mean-field shift and the broadening of the peak which can be calculated using a simple Fermi Golden Rule argument. 
The broad structure corresponds to the annihilation of a spin $\uparrow$ fermion in a classical gas of dimers. One is left with a single spin$\downarrow$ fermion in a gas of classical dimers. The kinetic energy of this fermion is $({\bf P}-{\bf k})^2/(2m)$ where ${\bf P}$ is the momentum of the dimer. This explains the shape of the curve as well as the existence of a threshold. 
All the necessary details are given in Appendix \ref{appendixBEClimit}. 

On the BCS side ($1/(k_F a)=-1$, $z=0.7$), the structure is simpler: a peak located around the kinetic energy $k^2/(2m)$ (also with a Lorentzian shape) with a finite broadening (see Fig.\ref{spectralfunction-1}). Again, this has a simple interpretation: the created spin $\uparrow$ fermion is going to scatter weakly on the spin $\downarrow$ fermions. As a consequence, there will be a mean field shift at the position of the peak. The width will be given by a Fermi Golden Rule calculation. More detailed arguments can be found in Appendix \ref{appendixBCSlimit}.

In the unitary limit ($1/(k_F a)=0$, $z=0.5$), a double-peak structure emerges at small momentum and it gradually evolves into one peak at large momentum. A small bump still exists near the main peak at the Fermi momentum, which makes it look very asymmetric. These characteristics cannot be interpreted simply in the same picture as on the BEC or BCS side. Coming from the BEC side, the bound state will disappear. Thus the corresponding left feature should disappear in the unitary limit and there should exist only one peak on the BCS side. Mathematically the double peak structure results from the real part of the self-energy since the second term of the imaginary part in Eq.(\ref{eqF_1}) vanishes and the remaining term is fairly smooth in the unitary limit. So we attribute this structure to the strong scattering between particles in the unitary limit and three-body physics or more-body physics may play a significant role in the structure of the spectral function.
\subsection{High negative frequency, high wave vector behavior and Tan's Contact}
\begin{figure}[h]
\subfigure[\label{fig1BEC}]{\includegraphics[width=0.65\linewidth]{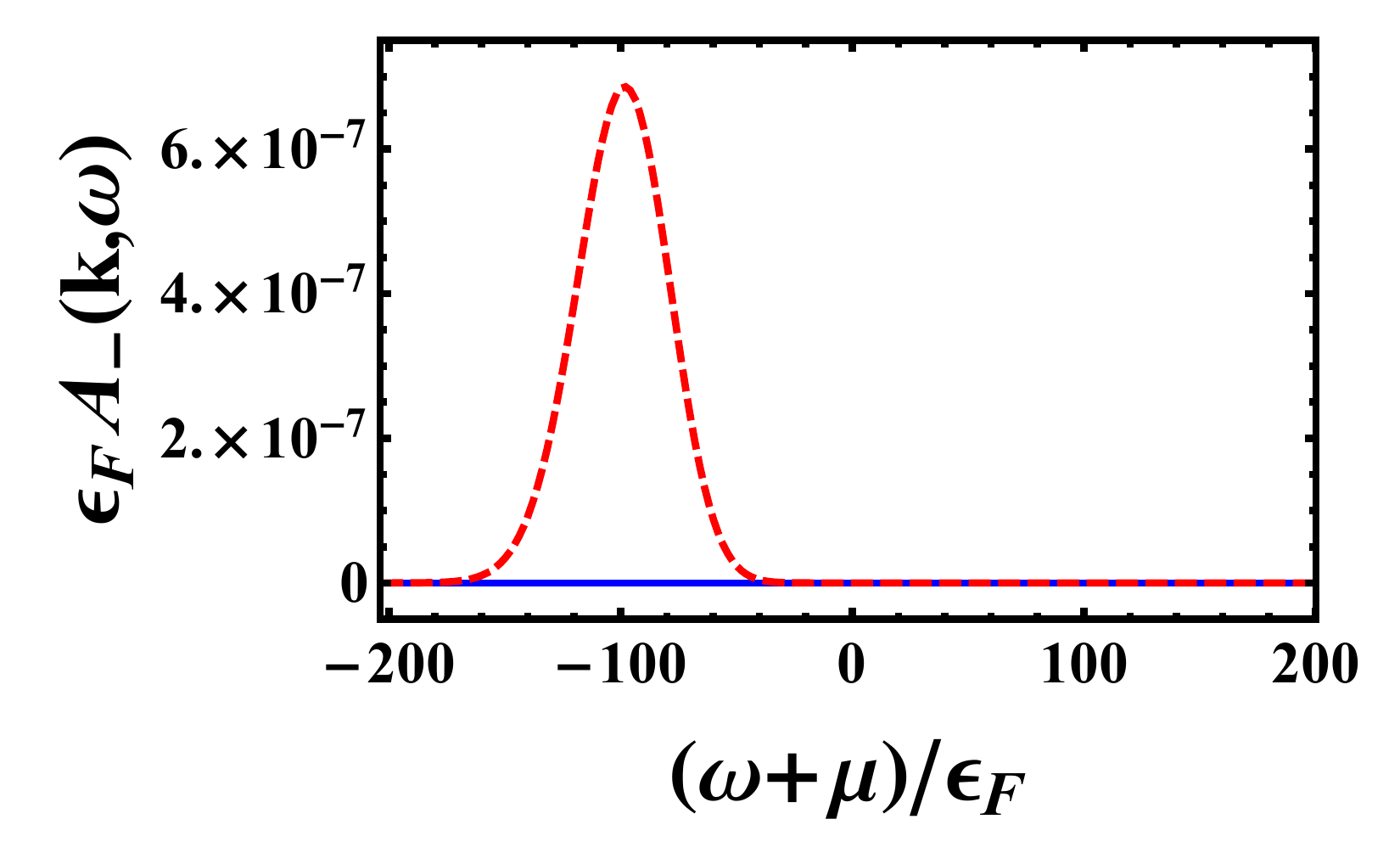}}\\
\subfigure[\label{fig1UL}]{\includegraphics[width=0.65\linewidth]{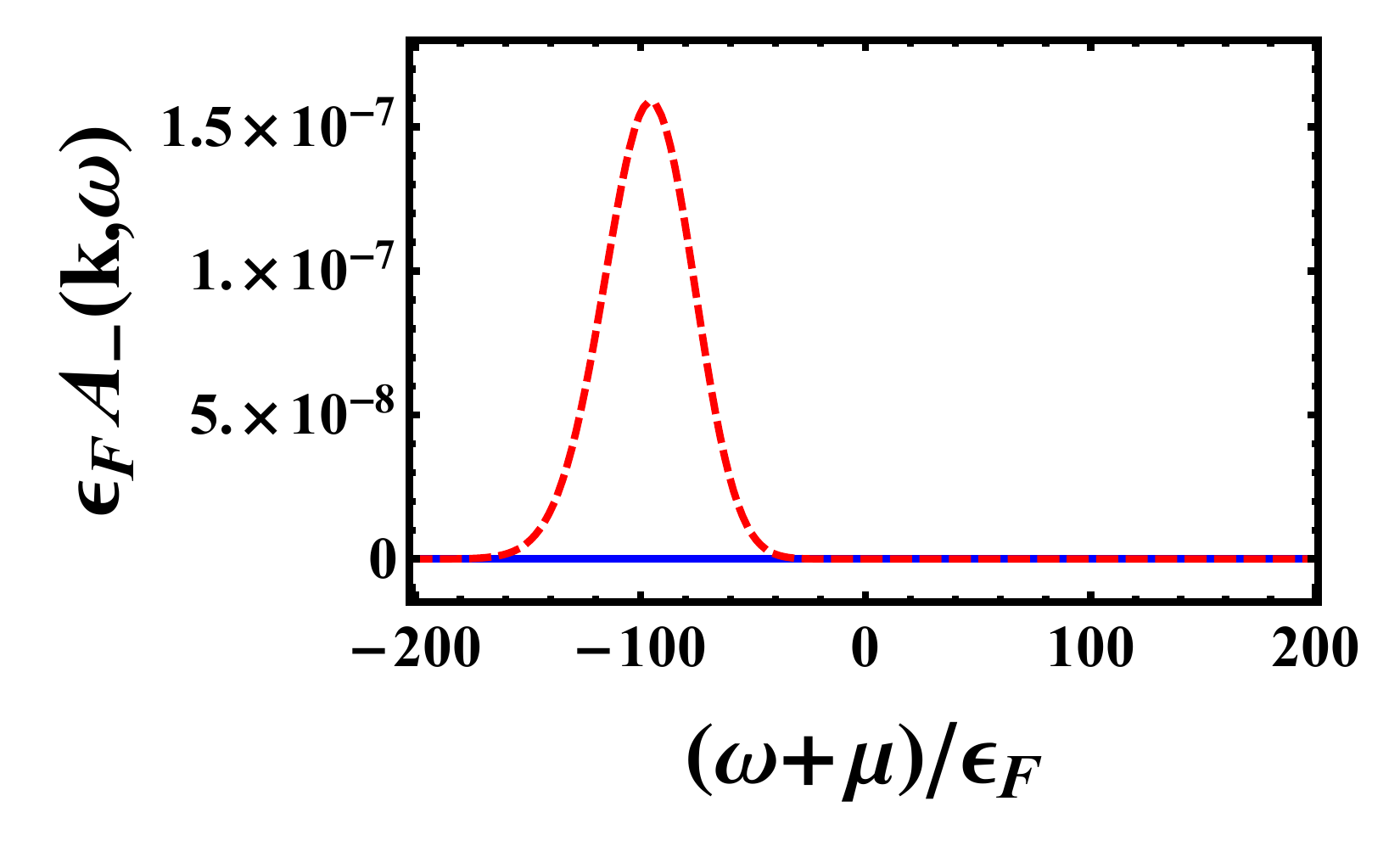}}\\
\subfigure[\label{fig1BCS}]{\includegraphics[width=0.65\linewidth]{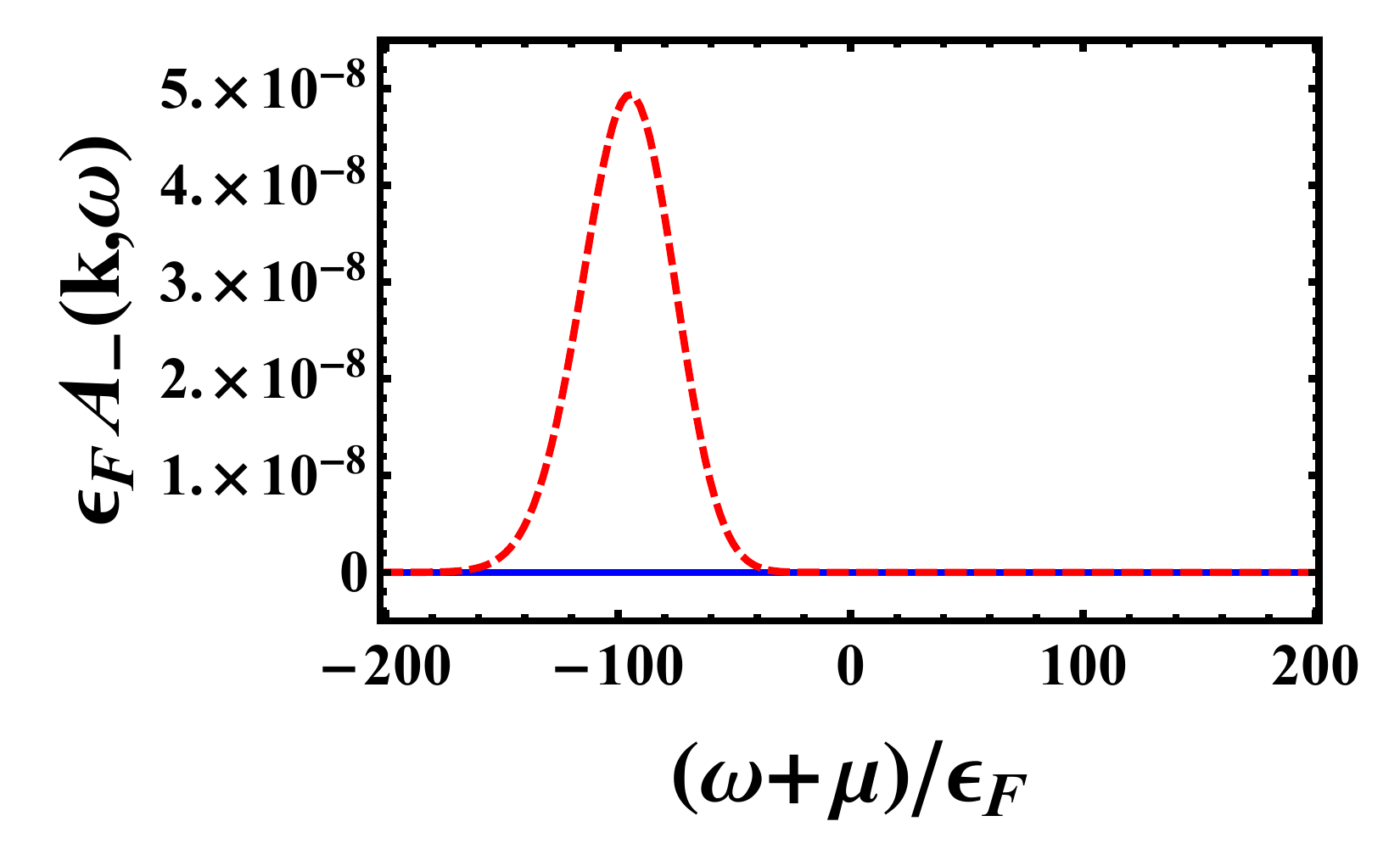}}
\caption{
The occupied spectral function $\varepsilon_F A_{-}(k,\omega)$ in a typical high wavector $k$ limit.  $T/T_F=1$ and $k/k_F=10$. (a): $1/(k_F a)=1$ (BEC regime), (b): $1/(k_F a)=0$  (Unitary Limit), (c): $1/(k_F a)=-1$ (BCS regime). The contributions with the self-energy of order $z$ is much smaller (blue continuous line) than the contribution of order $z^2$ (red dashed line).}
\end{figure}

For finite $\omega$ and $k$, the dominant contribution is the first term in Eq.(\ref{eqsigma_1}). It is of order $z$. However, if one is interested in evaluating the Tan's contact, one should consider the high $k$ limit. In this limit, as noted by the authors of Refs.\cite{praalzettocombescotleyronas,praschneiderranderia}, the dominant contribution to the spectral function or the imaginary part of the retarded self-energy comes from the region $\omega\simeq -k^2/(2m)$. However, in this limit, this is the imaginary part of the second term in Eq.(\ref{eqsigma_1}) which is dominant. Indeed, one finds easily, using Eqs.(\ref{eqF_1}) and (\ref{eqH_1}) that $\Im[F_1(k,\omega+\mu+i 0^+)]=e^{\beta(\omega+\mu)}\Im[H_1(k,\omega+\mu+i 0^+)]$. In the limit where $\beta\omega\to -\infty$ and
$|\omega|\gg |\mu|$, we find that the term of order $z^2$ is dominant. 
Therefore, the occupied spectral function $A_{-}(k,\omega)=f(\omega)A(k,\omega)$ ($f(\omega)=(e^{\omega/T}+1)^{-1}$ is the Fermi-Dirac distribution) in this high $k$ limit is dominated by the term of order $z^2$. This is shown in Figs.\ref{fig1BEC},\ref{fig1UL} and \ref{fig1BCS}. 
This has as a consequence that the Tan's contact is given by the apparently non-dominant term. 
In order to calculate the fermion occupation number $n_k$, one needs to integrate on the frequency the spectral function times the Fermi-Dirac distribution. In the high $k$ limit, the spectral function dominant contribution is essentially a Gaussian centered around $-k^2/(2m)$ and a width $(m T)^{-1/2} k$. This is explained in Appendix \ref{appendix high k}. The frequency integration is easily done and one finds at this order
\bea
n_k&\sim&\frac{C^{(2)}}{k^4}
\eea
The Tan's contact $C^{(2)}$ is given by
\beq
C^{(2)}=z^2 \frac{m^{7/2}T^{3/2}}{\pi^{3/2}}
[
\int_0^{+\infty} \ddx e^{-\beta x}\rho_2(x)+\Theta(a^{-1})Z_m e^{\beta E_b}
]
\label{eqC2}
\eeq
where we recognize the contributions coming from the continuum of two-body scattering states and from the dimer. It is easy to show that one recovers the known result. For instance, at the unitary limit, we find $C^{(2)}=z^2 m^2 T^2 4/\pi$ \cite{c2ybb}.

\begin{figure}[h]
\subfigure[\label{fignk1}]{\includegraphics[width=0.85\linewidth]{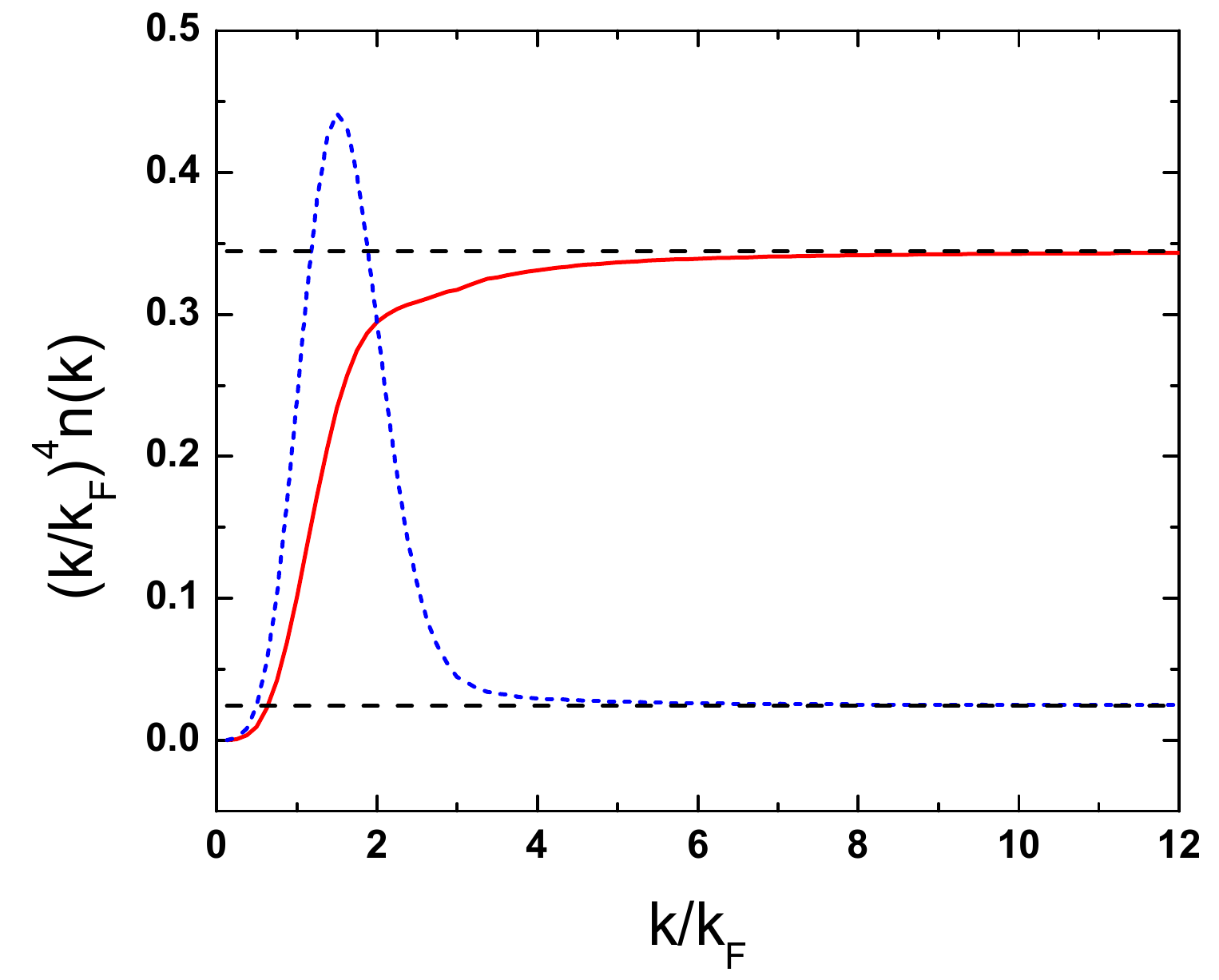}}\\
\subfigure[\label{fignk2}]{\includegraphics[width=0.85\linewidth]{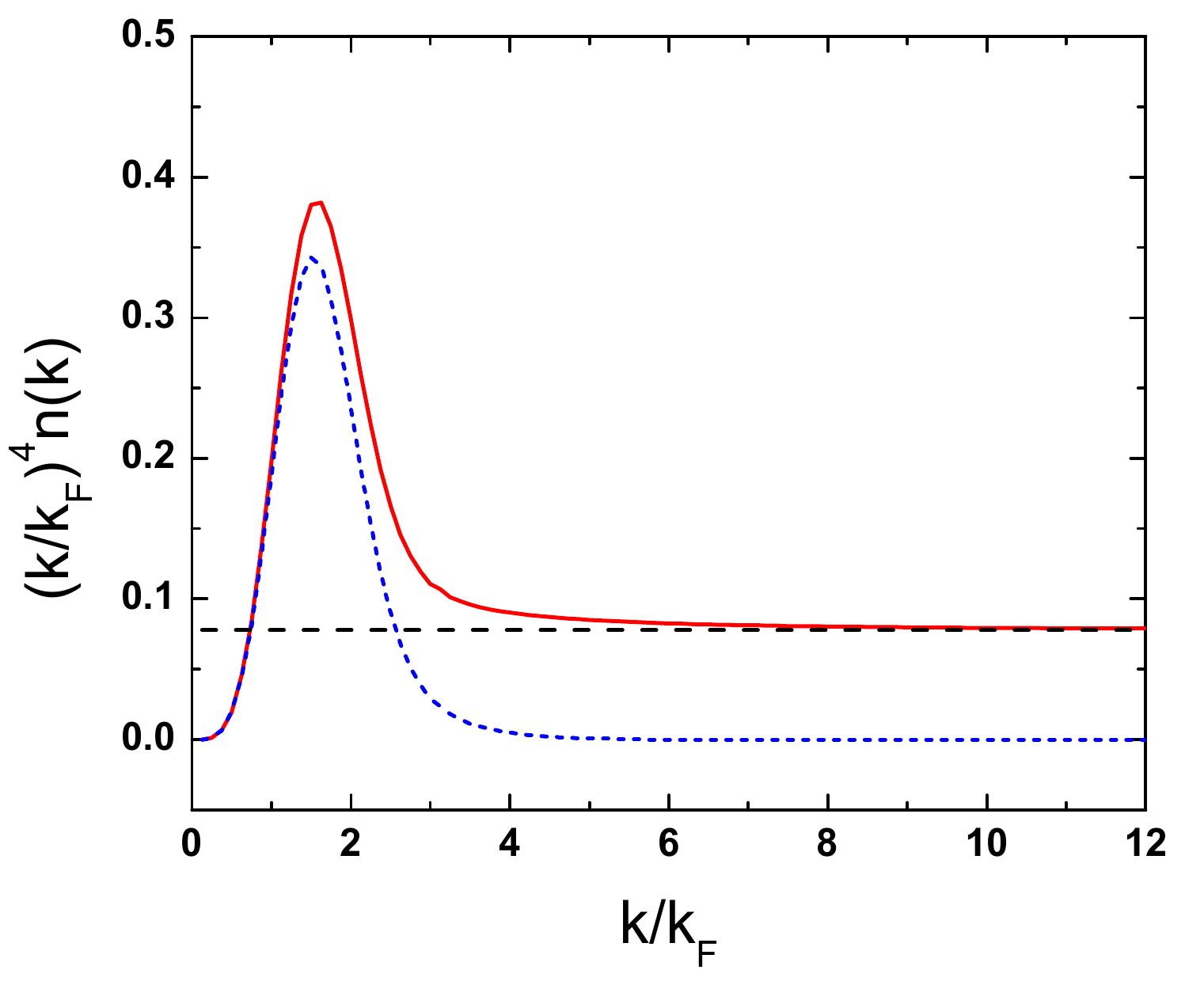}}
\caption{Momentum distribution $(k/k_F)^4 n_k$ for $T=T_F$. (a) : $1/(k_F a)=1$ (red solid line),  $1/(k_F a)=-1$ (blue dashed line). (b) : $1/(k_F a)=0$. The asymptotic values, given by Tan's contact at order $z^2$ (see Eq.(\ref{eqC2})) are shown as horizontal long dashed lines. For $1/(k_F a)=0$, the momentum distribution calculated with the term of order $z^2$ in the self-energy converges to Tan's contact (red full line). In contrast, without the term of order $z^2$ in the self-energy (blue dashed line), one does not recover Tan's contact.
}
\end{figure}

The occupation number $n_k$ is obtained by integrating on frequency the occupied spectral function $A_{-}(k,\omega)$. The results for the BEC regime, Unitary Limit and BCS regime are shown in Fig.\ref{fignk1} and \ref{fignk2}. We emphasize that it is necessary to include the self-energy up to order $z^2$ in order to get the correct asymptotic behavior $n_k\sim C^{(2)}/k^4$ for large $k$. We will see in section\ref{numUL}, that the other contributions of order $z^2$ do not contribute to the contact.

\section{Contact and pair-correlation functions including $3$-body correlations (order $z^3$)}\label{contact3}
We define the pair correlation function $\mathcal{P}({\bf r})=-\langle \left(\Psi^{\dagger}_{\uparrow}\Psi^{\dagger}_{\downarrow}\right)({\bf r},0)\left(\Psi_{\uparrow}\Psi_{\downarrow}\right)({\bf 0},0)\rangle$. As it was shown in \cite{kvhfwcontact}, the regular quantity for a short range interaction with a bare coupling constant $g_0$ is
$F({\bf r})=g_0^2 \mathcal{P}({\bf r})$. In the limit $g_0\to 0^{-}$, this quantity is related to the two-particle vertex function $\Gamma({\bf P},\tau)$ (sometimes called a "pair propagator") through the equation
$$
F({\bf r})=-\dP e^{-i{\bf P}\cdot{\bf r}}\Gamma({\bf P},\tau=0^{-})
$$
It turns out that it is more convenient to use the $\beta$-periodicity of $\Gamma({\bf P},\tau)$. We have
\bea
F({\bf r})&=&-\dP e^{-i{\bf P}\cdot{\bf r}}\Gamma({\bf P},\tau=\beta^{-})\label{eqdefF}
\eea
The contact $C$ is also related to the vertex function \cite{kvhfwcontact}
\bea
C&=&-m^2 \Gamma({\bf r}={\bf 0},\beta^-)
\eea
or equivalently, using Eq.(\ref{eqdefF}), $C=m^2 F({\bf r}={\bf 0})$.
\subsection{Order $z^2$: two-body correlations}
\begin{figure}[h]
\begin{center}
\includegraphics[width=0.5\linewidth]{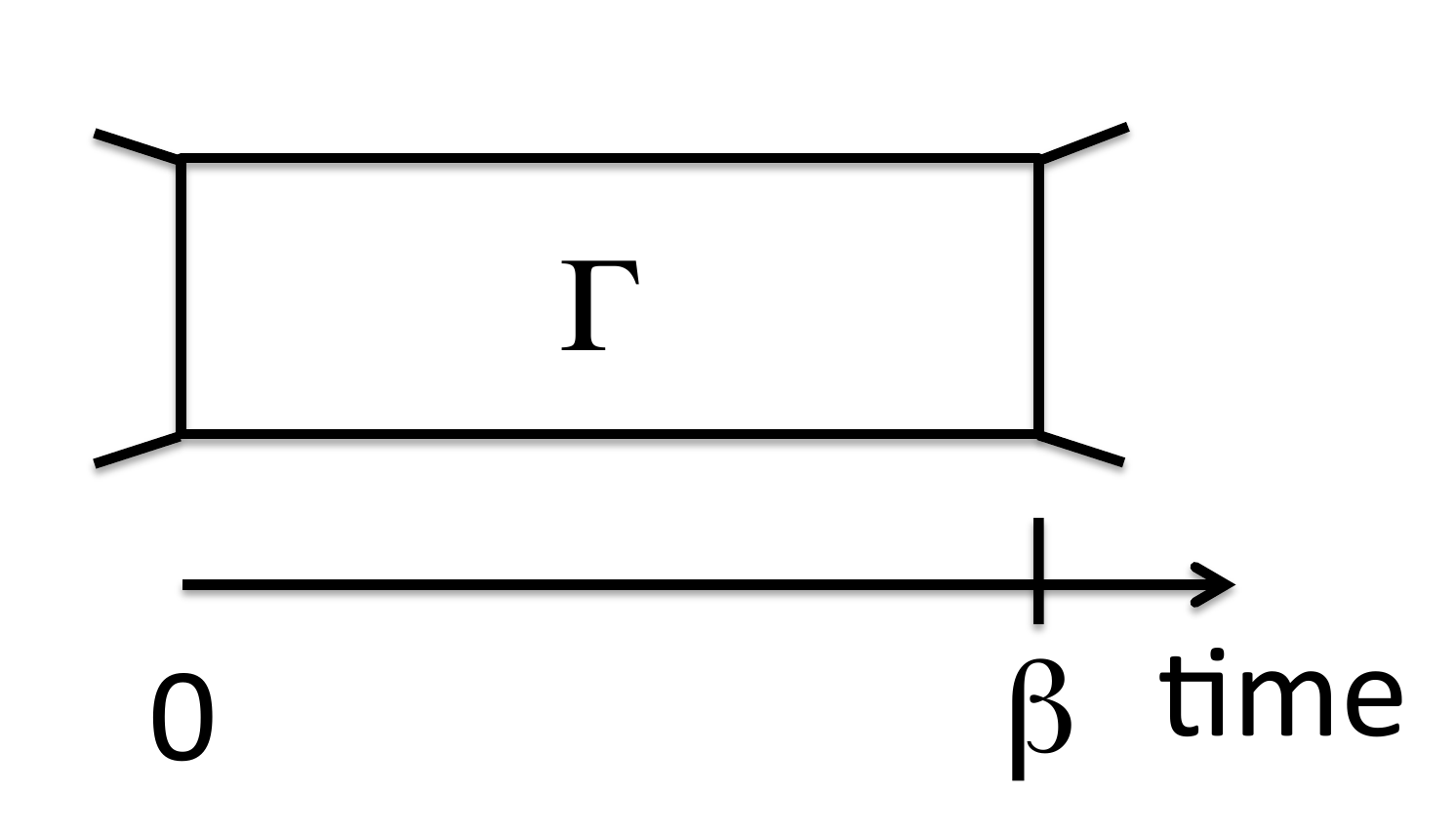}
\caption{Diagrammatic representation of $\Gamma({\bf P},\beta^-)$.}
\label{figdefGamma}
\end{center}
\end{figure}

In diagrammatic language, $\Gamma({\bf P},\beta^-)$ is represented in Fig.\ref{figdefGamma}. According to \cite{pravirial} we see directly that it contains at least a factor $e^{2\beta\mu}=z^2$ corresponding to the global time dependance of pair of fields. Indeed, in the Interaction representation each annihilation (respectively creation) operator $c_{{\bf k}}(\tau)=\exp(-\tau(\varepsilon_k-\mu))$ (respectively 
$c^{\dagger}_{{\bf k}}(\tau)=\exp(\tau(\varepsilon_k-\mu))$) has  a $\exp(\mu\tau)$ (resp. a $\exp(-\mu\tau)$) prefactor. We come to the first conclusion that there is no such exponential dependance for close loops, since they cancel. The second conclusion is that when a pair of particle is created at initial time $\tau=0$ and annihilated at time $\tau=\beta$ as is shown in Fig.\ref{figdefGamma} for the $2$-particle vertex, we get a factor $\exp(\beta\mu)^2=z^2$.
\begin{figure}[h]
\begin{center}
\includegraphics[width=0.5\linewidth]{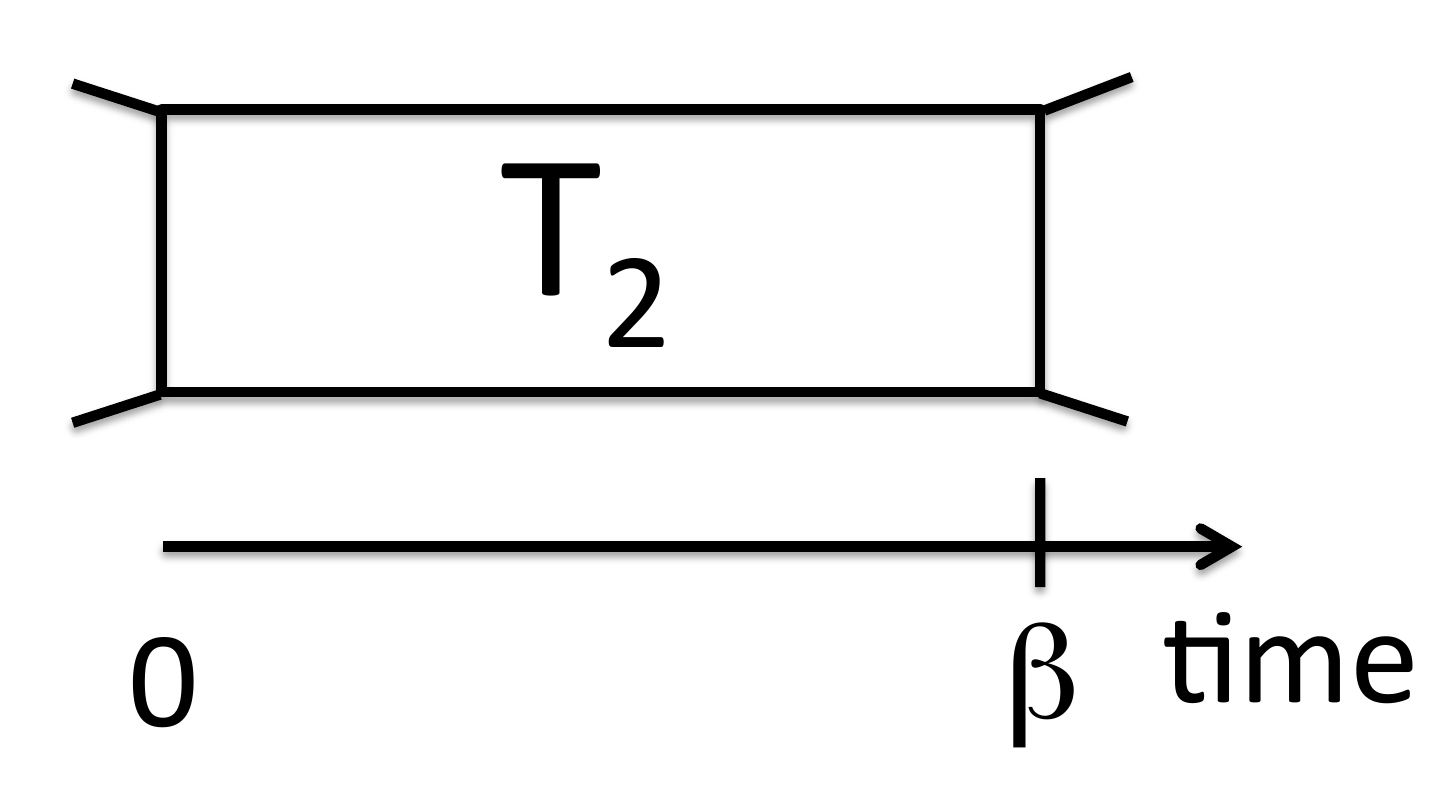}
\caption{The $2$-particle vertex $\Gamma({\bf P},\beta^-)$ at lowest order in the fugacity $z$.}
\label{figT2}
\end{center}
\end{figure}
Therefore, at lowest order, we can take the diagrams for $\Gamma$ in vacuum, which is $T_2({\bf P},\beta^-)$, see Fig.\ref{figT2}. In the diagrammatic expansion this is the sum of all the "ladder diagrams". We find
$\Gamma({\bf P},\beta^-)=z^2\, T_2({\bf P},\beta^-)+O(z^3)$. At order $z^2$, we find, using Eqs.(\ref{eqT_2_tau}) and (\ref{eqdefF})
\bea
F^{(2)}({\bf r})&=&z^2 \,\dP e^{-i{\bf P}\cdot{\bf r}}
e^{-\beta\frac{P^2}{4 m}}\nonumber\\
&\times&[\Theta(a^{-1})Z_m e^{\beta E_b}+\int_0^{+\infty}\ddx e^{-\beta x}\rho_2(x)]
\eea
We can perform the integral on ${\bf P}$ and we get
\bea
F^{(2)}({\bf r})&=&\frac{C^{(2)}}{m^2}e^{-\frac{m r^2}{\beta}}
\eea
where $C^{(2)}$ is the lowest-order value of the contact given in Eq.(\ref{eqC2}).
\subsection{Order $z^3$ : three-body correlations}\label{seccontact3}
\begin{figure}[h]
\subfigure[\label{figGamma2_T3a}]{\includegraphics[width=0.65\linewidth]{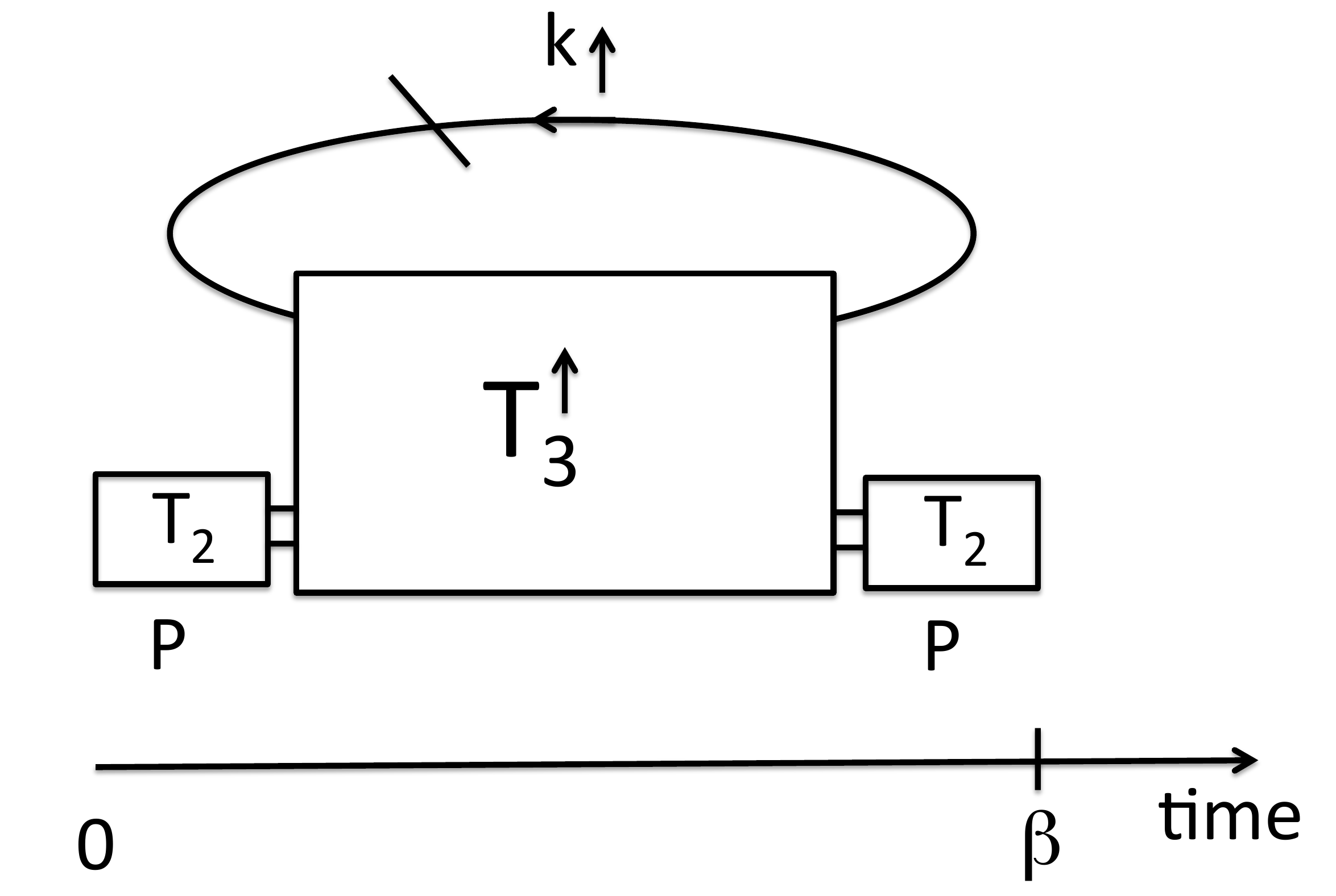}}\\
\subfigure[\label{figGamma2_T3b}]{\includegraphics[width=0.65\linewidth]{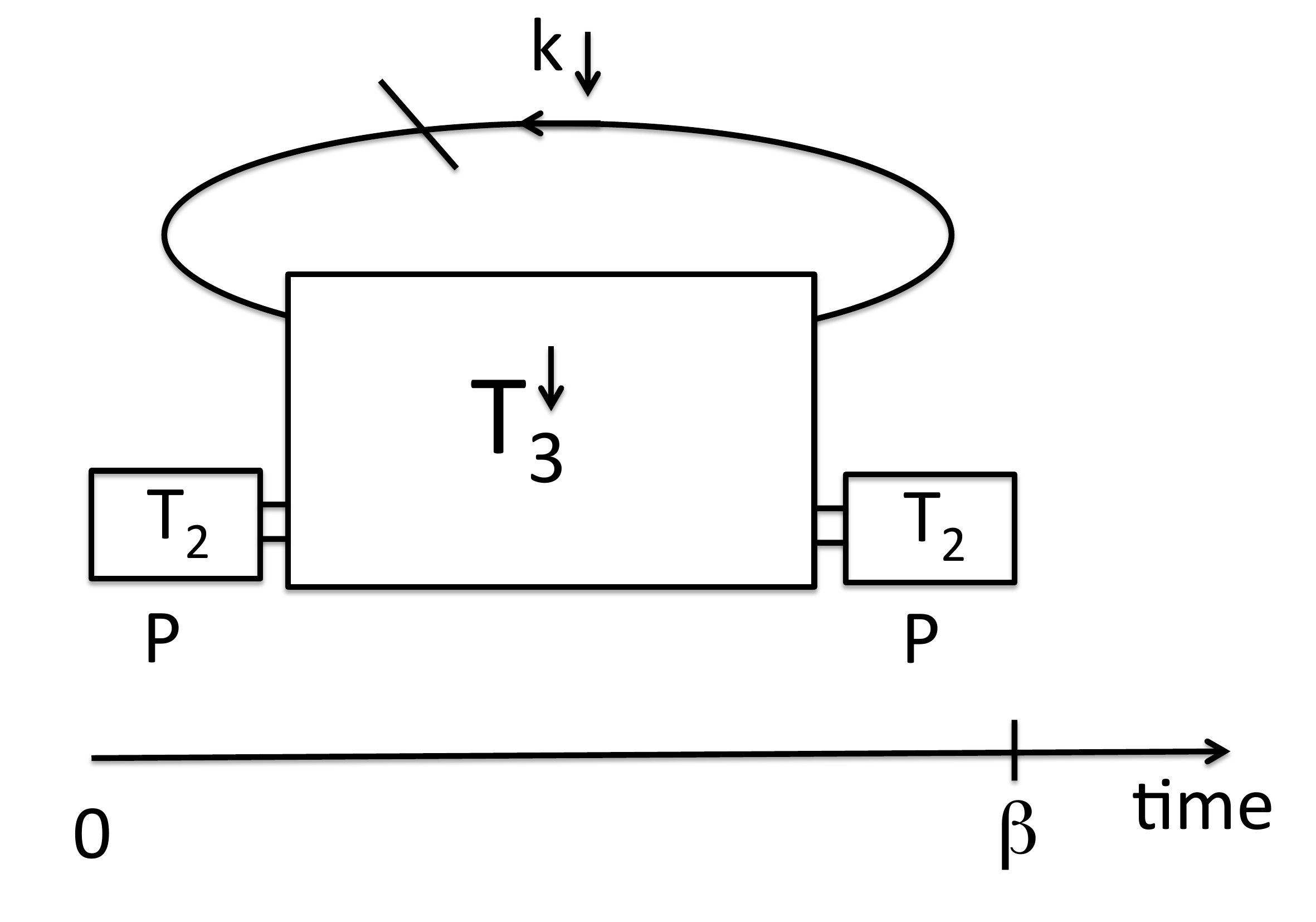}}
\caption{The two diagrams contributing to $\Gamma^{(3)}({\bf P},\beta^-)$, and hence to $c_3$, the virial expansion term of Tan's contact coefficient of order $z^3$ (see Eq.(\ref{eqdefc3})).
}
\end{figure}

\begin{figure}[h]
\begin{center}
\includegraphics[width=0.9\linewidth]{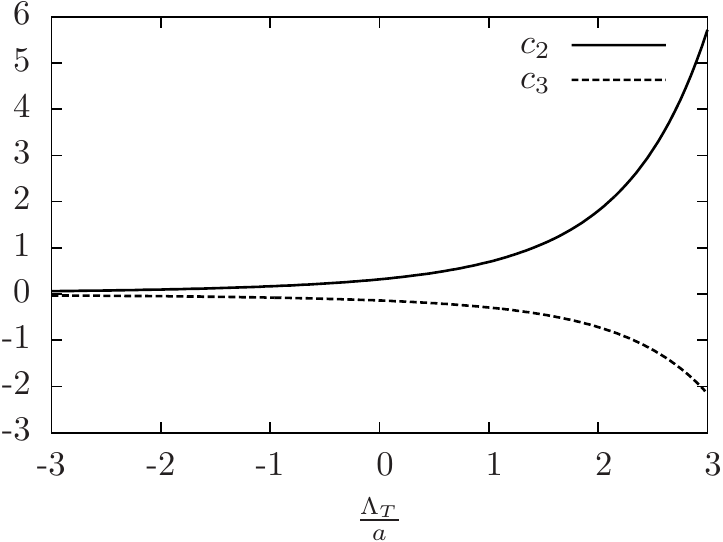}
\caption{The coefficients $c_2$ (continuous line) and $c_3$ (dashed line) as defined in Eq.(\ref{eqdefc3}) for the virial expansion of Tan's contact, in the whole BEC-BCS crossover.}
\label{figc2c3}
\end{center}
\end{figure}

At next order, we must find all the diagrams with one slashed line. They are shown in Fig.\ref{figGamma2_T3a} and \ref{figGamma2_T3b}. 
They involve the $3-$body problem through $T_3$, the $3-$body $T$-matrix. $T_3$ can be easily obtained numerically, solving linear integral equations. This is explained in details in Appendix B of \cite{pravirial}. These equations have been first found  by Skorniakov and Ter-Martirosian in their seminal work on the $3$-body problem \cite{stm} and Ref.\cite{pra4par}.
The analytical expressions are given in Appendix \ref{appendixGamma2T_3}. From this we can deduce a virial expansion of the contact
\bea
C&=&\frac{16\pi^2}{\Lambda_T^4}\left(c_2 z^2+c_3 z^3+\cdots
\right)\label{eqdefc3}
\eea
where the thermal wavelength $\Lambda_T=\sqrt{2\pi \beta/m}$.
The coefficients $c_2$ (from Eq.(\ref{eqC2})) and $c_3$ are plotted as a function of the relevant dimensionless parameter $\Lambda_T/a$ in Fig.\ref{figc2c3}. From Tan's adiabatic theorem, we easily find that
$c_n=\partial b_n/\partial(\Lambda_T/a)$, where the $b_n$ are virial cumulants entering in the expansion of the equation of state. We have cross-checked the results obtained for $b_3$ in \cite{pravirial} and $c_3$ (this work) using this relation, finding a good agreement.

\section{Expressions for the self-energy including $3$-body correlations}\label{secsigma2}
We now show our results for the self-energy at second order, {\it i.e.} with a double slashed fermionic line or with two slashed lines. We give analytical results for $a^{-1}\leq 0$ and equal masses.
We performed numerical calculations for the spectral function at the unitary limit where $a^{-1}=0$.
\subsection{$\Sigma^{(2,1)}$}
\begin{figure}[h]
\begin{center}
\includegraphics[width=0.7\linewidth]{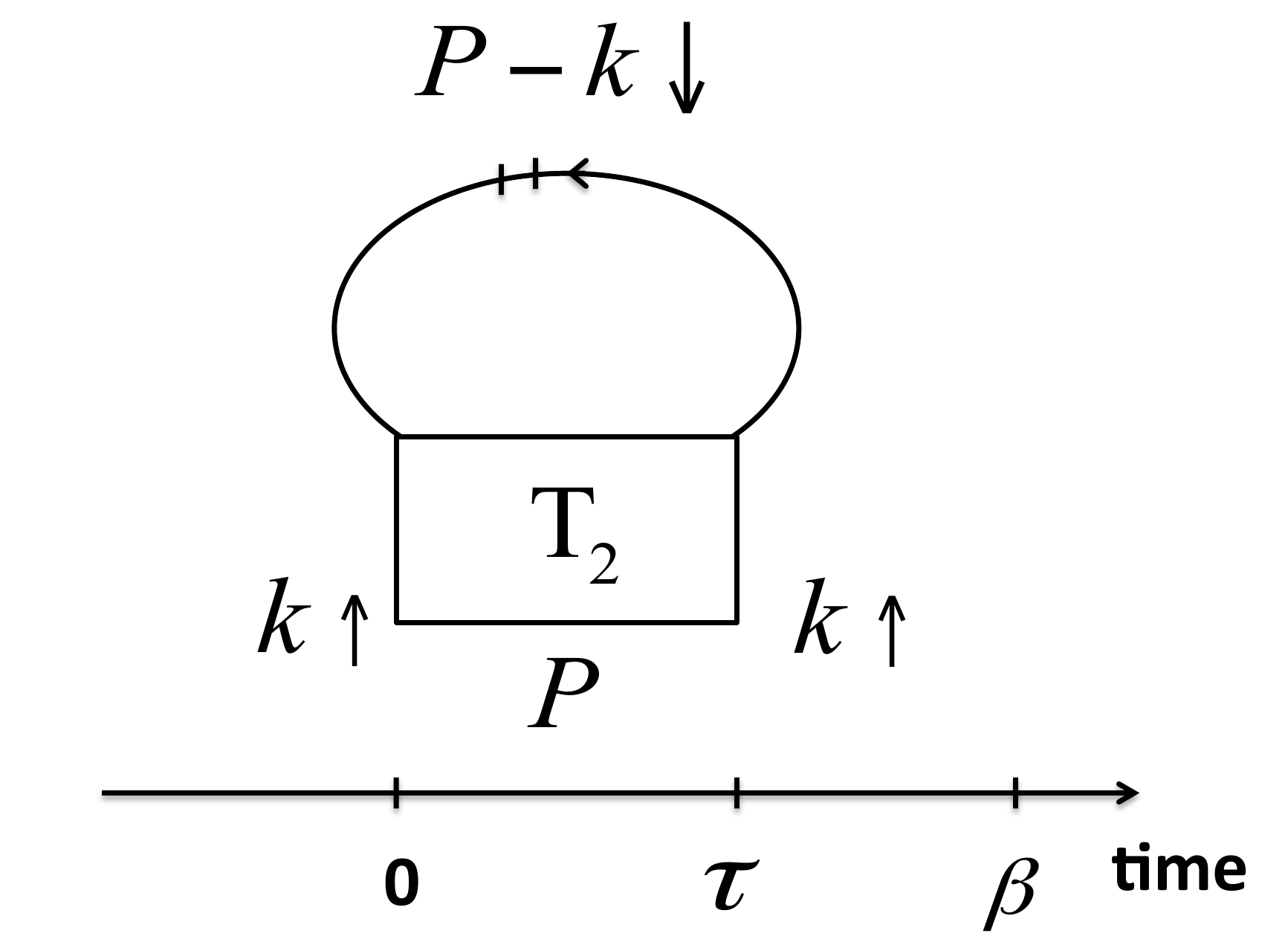}
\caption{The Feynman diagram contributing to $\Sigma^{(2)}$ containing one $G^{(0,2)}$ line (doubly slashed fermionic line).}
\label{figSig2_1}
\end{center}
\end{figure}

The only diagram with a double slashed line is shown in Fig.\ref{figSig2_1}, and is obtained from Fig.\ref{figsigma1} by changing the slashed line into a double slashed line.
The analytic expression is easily obtained from the expression of $\Sigma^{(1)}$ by multiplying by a factor $z=e^{\beta\mu}$ and changing $G^{(0,1)}$ into $G^{(0,2)}$, which amounts to multiplying the integrand in Eq.(\ref{eqsigma1_tau}) by $-e^{-\beta\varepsilon_{{\bf P-k}}}$. So we have the contribution to the retarded self-energy
\begin{equation}
\Sigma^{(2,1)}_R(k,\omega)=z^2\, F_{2,1}(k,\omega+\mu+i 0^+)+z^3 \,H_{2,1}(k,\omega+\mu+i 0^+)\label{eqsigma_21}
\end{equation}
 where
\begin{multline}
 F_{2,1}(k,E)=-\dP\int_0^{+\infty}\hspace{-2.mm} \ddx\frac{e^{-\beta\frac{({\bf P-k})^2}{m}}\rho_2(x)}{E-(\frac{P^2}{4m}+x-\frac{({\bf P-k})^2}{2 m})} \label{eqF21}
 \end{multline}
 \begin{multline}
 H_{2,1}(k,E)= -\dP\int_0^{+\infty}\hspace{-2.mm} \ddx\hspace{2.mm}\frac{e^{-\beta(x+\frac{({\bf P-k})^2}{2m}+\frac{P^2}{4m})}\rho_2(x)}{E-(\frac{P^2}{4m}+x-\frac{({\bf P-k})^2}{2 m})} \label{eqH21}
\end{multline}

 Following the analysis of Appendix \ref{appendix high k}, we find that this diagram does not contribute to the contact.

\subsection{$\Sigma^{(2,2)}$}
The diagrams with two $G^{(0,1)}$'s are shown in Figs.\ref{figsig23a}-\ref{figsig23f}. We first detail the calculation of
diagrams of Figs.\ref{figsig23a},\ref{figsig23b} and \ref{figsig23c}. The diagrams of Figs.\ref{figsig23d},\ref{figsig23e} and \ref{figsig23f} deserve a separate treatment.
They all involve the $3-$body problem through $T_3$, the $3-$body $T$-matrix.

\subsubsection{$\Sigma^{(2,2,a)}$}
The diagram of Fig.\ref{figsig23a} can be redrawn using the unfolding trick in a way similar to Fig.\ref{figGamma2_a_unfold}. In the time domain, we see that it is a convolution product of the form (we do not write the wave vector indices for simplicity)
\begin{multline}
e^{\mu\tau}e^{-(\beta-\tau)(\varepsilon+\varepsilon')}\int_{\mathcal{D}}dt_1 dt_2
e^{-\varepsilon' t_1}T_2(t_1)T_3^{\downarrow}(t_2)\\
\times e^{-\varepsilon'(\tau-t_1-t_2)}T_2(\tau-t_1-t_2)\nonumber
\end{multline}
where $\varepsilon^{\prime}=k'^2/(2m)$, $\varepsilon=({\bf P-k})^2/(2m)$ and $\mathcal{D}=\{t_1>0,t_2>0,\tau-t_1-t_2>0\}$. The double time integral can therefore be replaced by the inverse Laplace transform
\beq
e^{\mu\tau}e^{-(\beta-\tau)(\varepsilon+\varepsilon')}\ds
e^{-\tau s}T_2(s-\varepsilon')T_3^{\downarrow}(s)T_2(s-\varepsilon')\label{eqsiga}
\eeq
where $\gamma$ is such that the integrand is analytic for $\Re(s)<\gamma$ and $\mathcal{C}_{\gamma}$ is a Bromwich contour.
In order to get the contribution to the self-energy, we need to integrate on wave vectors. ${\bf k}'$ is the wave vector of the slashed spin down Green's function  and ${\bf P}$ is the wave vector of the two $T_2$ matrices.
As usual, we go to the center of mass reference frame, defining the total momentum ${\bf P_t}={\bf P}+{\bf k'}$ and relative motion atom-dimer momenta
${\bf p'}_1={\bf k'}-1/3 {\bf P_t}$. In the following, we will need the function $F({\bf p'}_1,{\bf p'}_2;s)=t_2(s-3/4 (p'_{1})^{2}/m)\,t_3^{\uparrow}({\bf p'}_1,{\bf p'}_2;s)\,t_2(s-3/4 (p'_{2})^2/m)$, and the associated spectral function 
$\rho_3({\bf p'}_1,{\bf p'}_2;x)\equiv -1/\pi \Im[F({\bf p'}_1,{\bf p'}_2;x+i\,0^+)]$.
The next step is to deform the contour integral in Eq.(\ref{eqsiga}) around the real axis, as it is done for the calculation of $\Sigma^{(1)}$ in Eq.(\ref{eqT_2_tau}). Then, we can take the Fourier transform $\int_{0}^{\beta} e^{i\omega_n\tau}\cdot$. The only dependence on $\omega_n$ then appears in a denominator of the form
$i\omega_n+\mu+\varepsilon+\varepsilon'-x$. We can safely perform the analytical continuation to the real axis in order to get the retarded self-energy
by replacing $i\omega_n$ by $\omega+i 0^{+}$.
We finally get
\begin{multline}
\Sigma^{(2,2,a)}_R(k,\omega)=z^2\, F_{2,2,a}(k,\omega+\mu+i 0^+)+\\
z^3 \,H_{2,2,a}(k,\omega+\mu+i 0^+)\label{eqsigma_221}
\end{multline} 
with a term of order $z^2$ and a term of order $z^3$.
The expressions for $F_{2,2,a}$ and $H_{2,2,a}$ are
\begin{multline}\label{eqsigmaF221}
F_{2,2,a}(k,E)=\dppp\,
\rho_{3}({\bf p'}_1,{\bf p'}_1;x)\\
\times\frac{e^{-\beta\mathcal{E}_a({\bf P}_t,{\bf p'}_1,{\bf k})}}
{E-x-\frac{P_{t}^2}{6m}+\mathcal{E}_a({\bf P}_t,{\bf p'}_1,{\bf k})}
\end{multline}
\begin{multline}
H_{2,2,a}(k,E)=\dppp\,
\rho_{3}({\bf p'}_1,{\bf p'}_1;x)\\
\times\frac{e^{-\beta(x+\frac{P_{t}^2}{6m})}}
{E-x-\frac{P_{t}^2}{6m}+\mathcal{E}_a({\bf P}_t,{\bf p'}_1,{\bf k})}
\end{multline}
where $\mathcal{E}_a({\bf P}_t,{\bf p'}_1,{\bf k})=\frac{({\bf P}_t/3+{\bf p'}_1)^2}{2m}+\frac{(2{\bf P}_t/3-{\bf p'}_1-{\bf k})^2}{2m}$.
$F_{2,2,a}$ and $H_{2,2,a}$ have the same structure as $F_1$ and $H_1$ respectively. The energy denominator corresponds to the energy of intermediate states consisting of $3$ particles (relative motion energy $x$, center of mass kinetic energy $\frac{P_{t}^2}{6m}$) and  of two holes (energy $-\frac{({\bf P}_t/3+{\bf p'}_1)^2}{2m}$ and $-\frac{(2{\bf P}_t/3-{\bf p'}_1-{\bf k})^2}{2m}$).

\subsubsection{$\Sigma^{(2,2,b)}$}
The self-energy of diagram Fig.\ref{figsig23b}, is denoted $\Sigma^{(2,2,b)}$. In the case of equal masses, it is equal to the contribution of diagram of Fig.\ref{figsig23a}, and we have
\bea
\Sigma_R^{(2,2,b)}(k,\omega)&=&\Sigma_R^{(2,2,a)}(k,\omega)
\eea
\subsubsection{$\Sigma^{(2,2,c)}$}
\begin{figure}[h]
\begin{center}
\subfigure[\label{figsig23a}]{\includegraphics[width=0.49\linewidth]{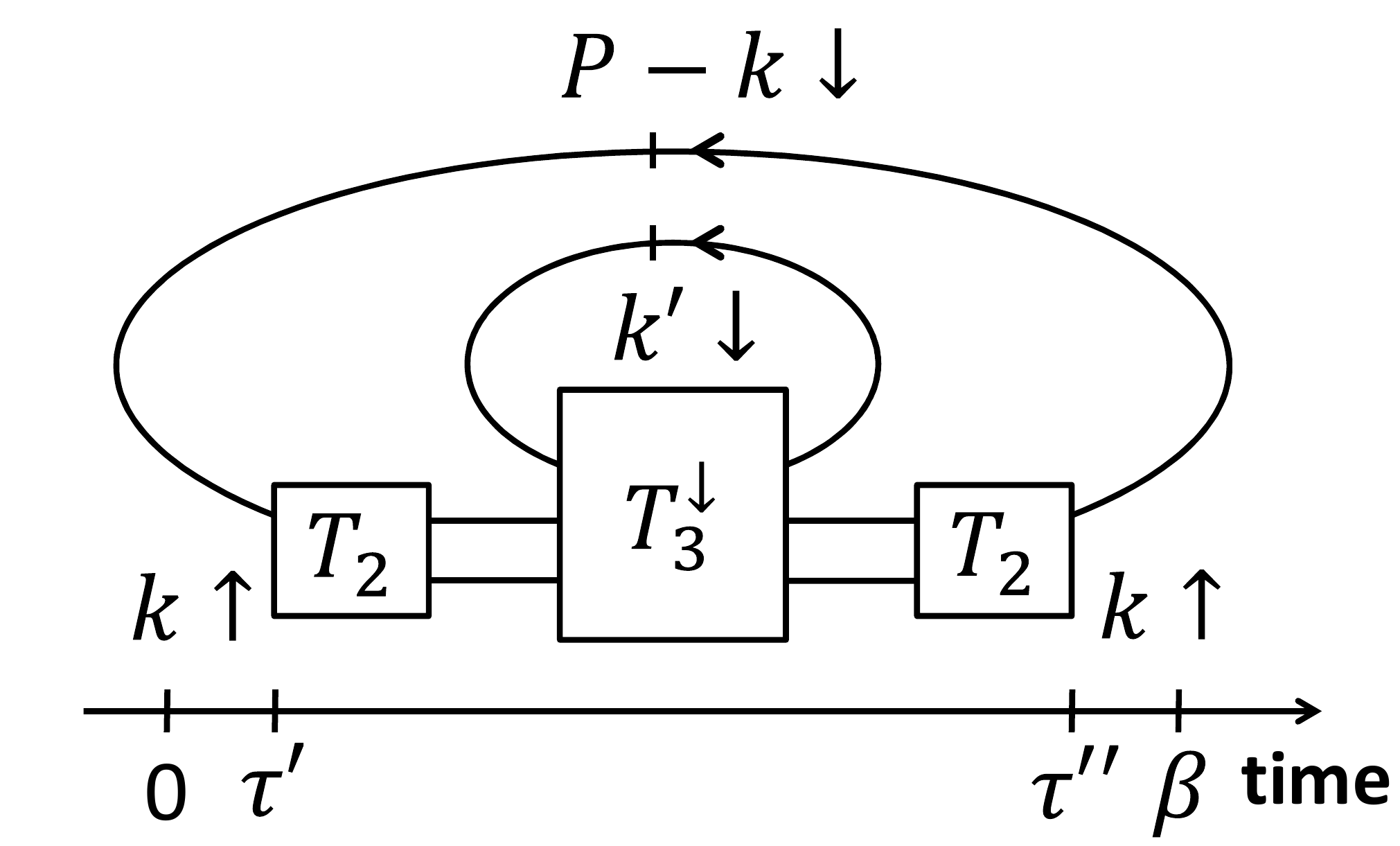}}
\subfigure[\label{figsig23b}]{\includegraphics[width=0.49\linewidth]{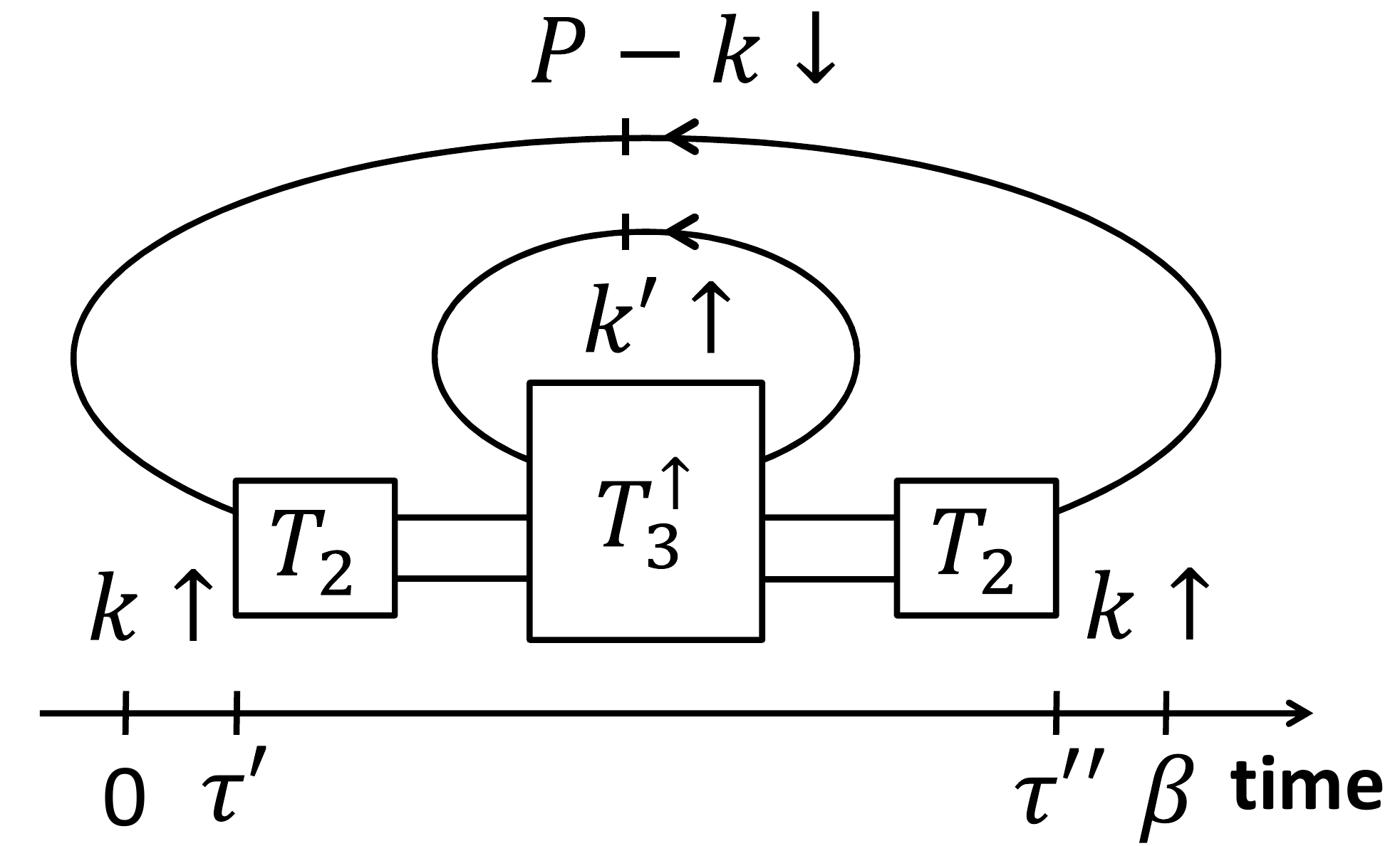}}\\
\subfigure[\label{figsig23c}]{\includegraphics[width=0.48\linewidth]{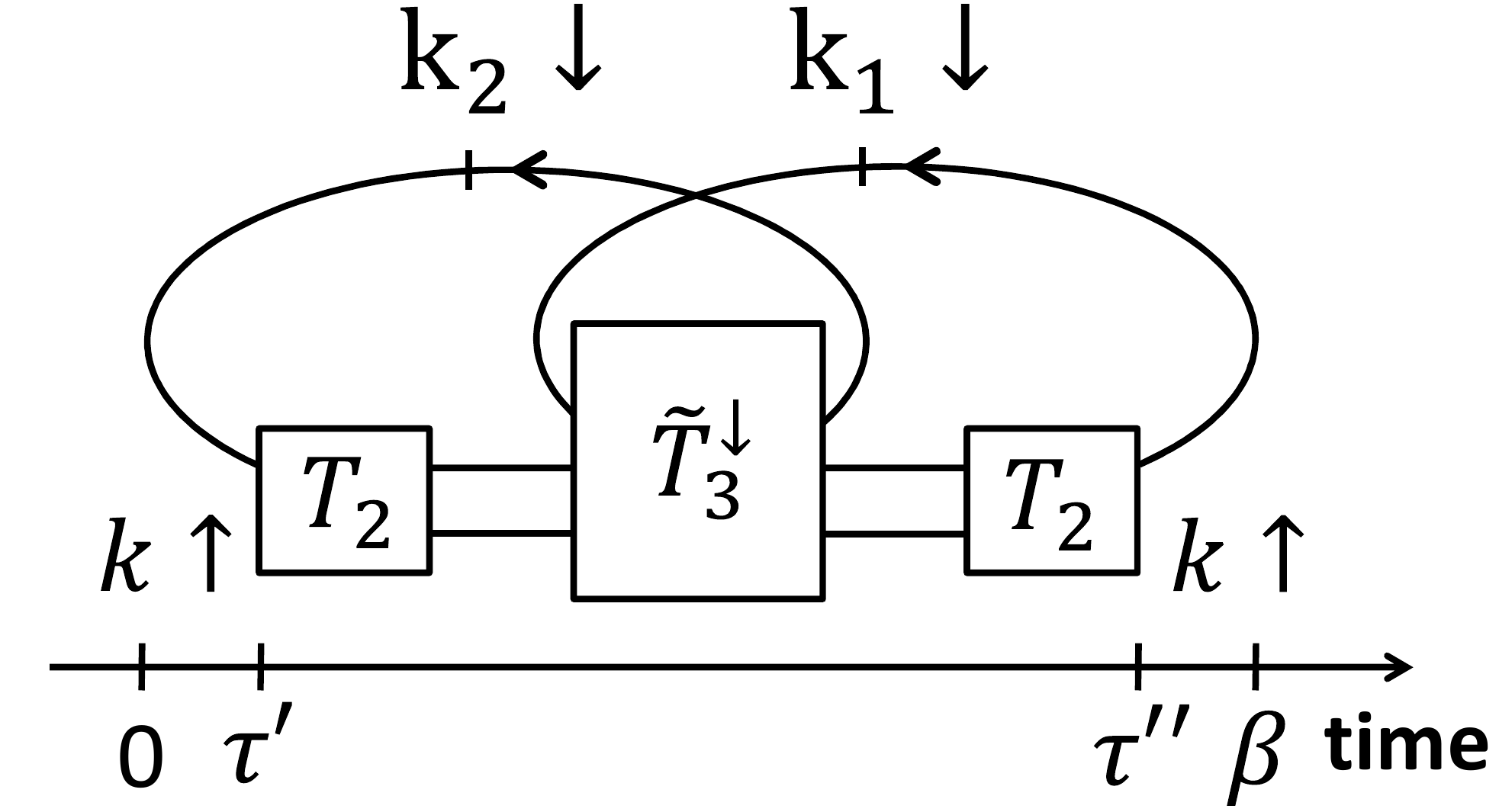}}
\subfigure[\label{figsig23d}]{\includegraphics[width=0.48\linewidth]{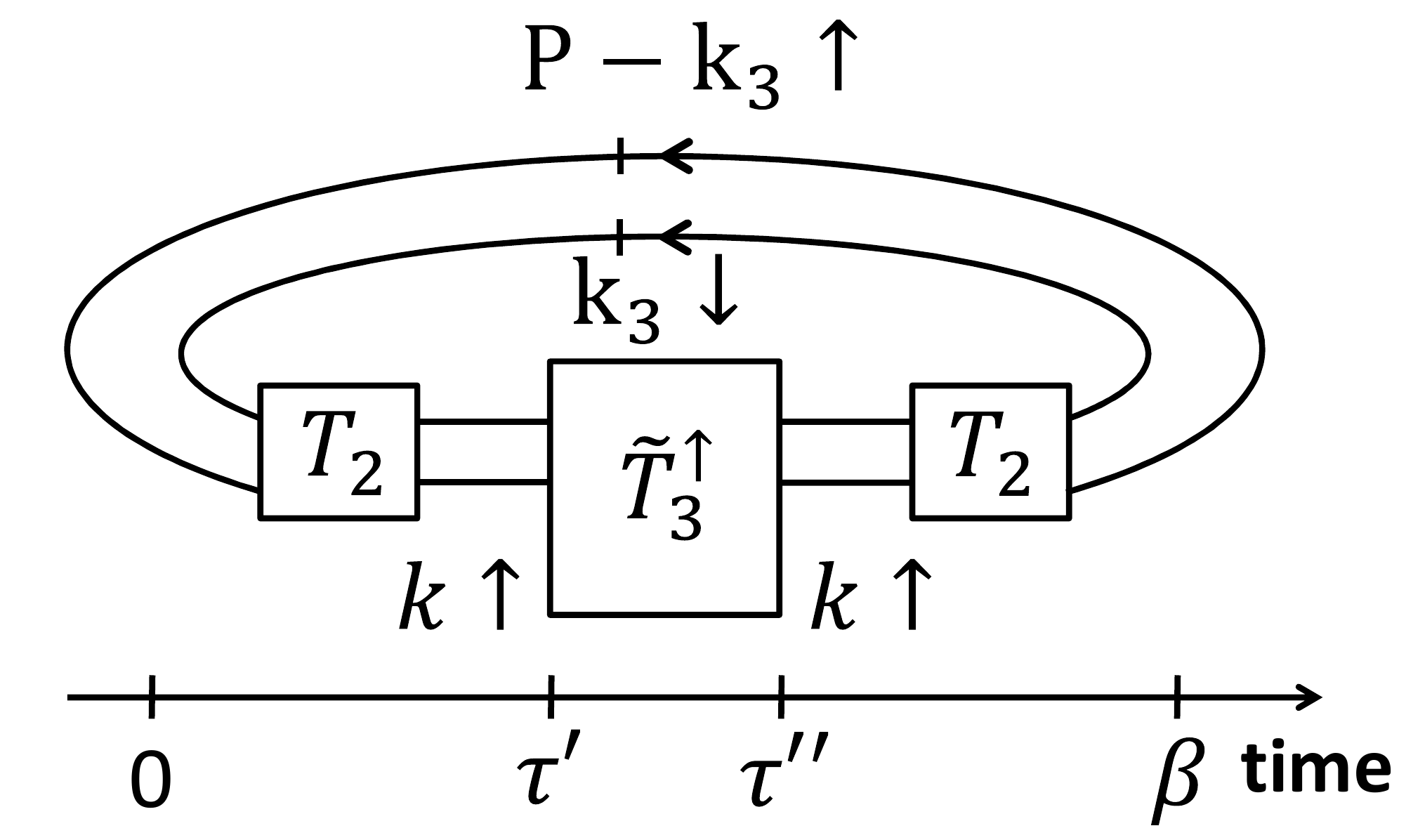}}\\
\subfigure[\label{figsig23e}]{\includegraphics[width=0.48\linewidth]{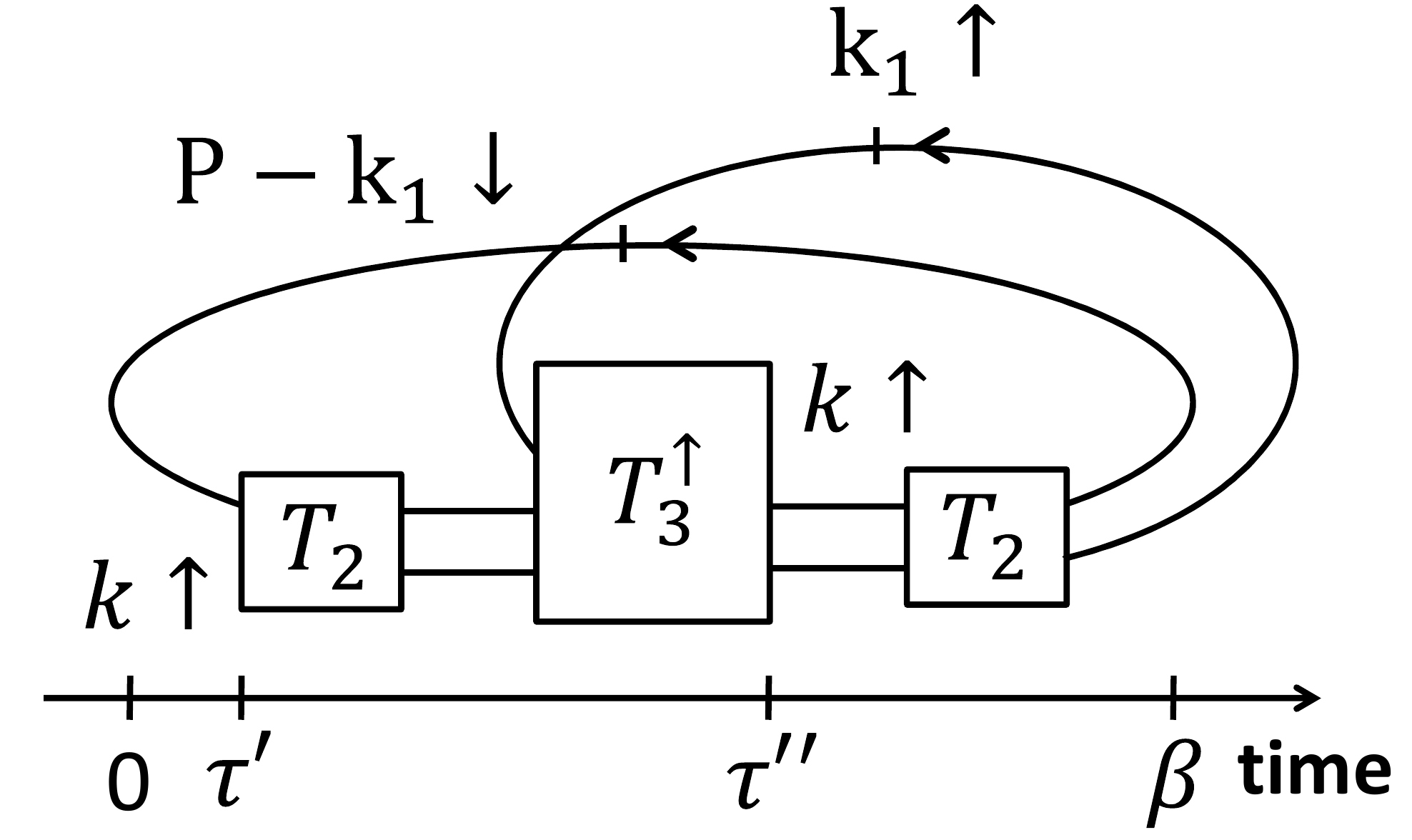}}
\subfigure[\label{figsig23f}]{\includegraphics[width=0.48\linewidth]{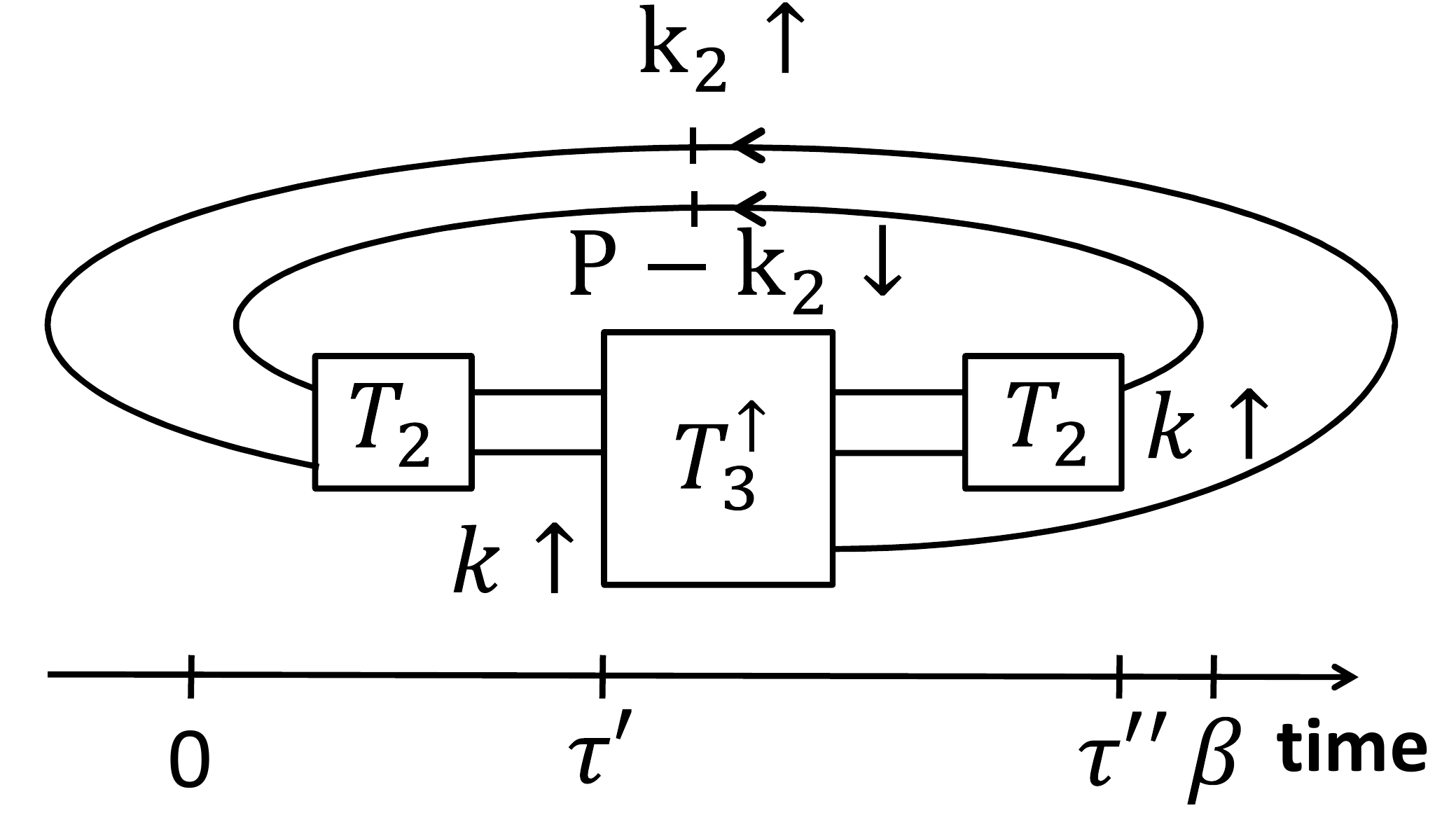}}
\caption{The $6$ diagrams contributing to $\Sigma^{(2)}(k,\tau"-\tau')$, including the $3$-body problem.}
\label{figsig23af}\end{center}
\end{figure}

The calculation of the diagram of Fig.\ref{figsig23c} goes along the same line of reasoning as for $\Sigma^{(2,2,a)}$. Due to the exchange of identical fermions, we find a global minus sign. Moreover, a careful inspection shows that the diagram corresponding to the Born approximation  for the $3$-body problem is not a self-energy diagram (not $1$-particle irreducible) and therefore we should substract it. We denote by $\tilde{T}_3$ the new $3$-particle $T$-matrix with the subtracted Born approximation (see Fig.\ref{figsig23c}) and $\tilde{\rho}_3$ the corresponding spectral function. 
For the on-shell $3$-body $T-$matrix $t_3$, we have 
\bea
\tilde{t}_3^{\uparrow}({\bf p}'_1,{\bf p}'_2;s)&=&
t_3^{\uparrow}({\bf p}'_1,{\bf p}'_2;s)-
\frac{1}
{
\frac{(p'_1)^2+(p'_2)^2}{m}+
\frac{{\bf p'}_1\cdot{\bf p'}_2}{m}-s
}\nonumber\\
\eea

We have 
\begin{equation}
\Sigma_R^{(2,2,c)}(k,\omega)=z^2 F_{2,2,c}(k,\omega+\mu+i 0^+)+z^3 H_{2,2,c}(k,\omega+\mu+i 0^+)
\end{equation}
The expressions for $F_{2,2,c}$ and $H_{2,2,c}$ are
\bea
F_{2,2,c}(k,E)&=&-3^3\dpp\,
\tilde{\rho}_{3}({\bf p'}_1,{\bf p'}_2;x)\nonumber\\
&\times&\frac{e^{-\beta\mathcal{E}_c({\bf p'}_1,{\bf p'}_2,{\bf k})}}
{E-x-\frac{P_{t}^2}{6m}+\mathcal{E}_c({\bf p'}_1,{\bf p'}_2,{\bf k})}\label{eqf223}
\eea
\bea
H_{2,2,c}(k,E)&=&-3^3\dpp\,
\tilde{\rho}_{3}({\bf p'}_1,{\bf p'}_2;x)\nonumber\\
&\times&\frac{e^{-\beta(x+\frac{P_{t}^2}{6m})}}
{E-x-\frac{P_{t}^2}{6m}+\mathcal{E}_c({\bf p'}_1,{\bf p'}_2,{\bf k})}\label{eqh223}
\eea
where ${\bf P}_t=3({\bf p'}_1+{\bf p'}_2+{\bf k})$, $\mathcal{E}_c({\bf p'}_1,{\bf p'}_2,{\bf k})=\frac{(2{\bf p'}_1+{\bf p'}_2+{\bf k})^2}{2m}+\frac{({\bf p'}_1+2{\bf p'}_2+{\bf k})^2}{2m}$.
The factor $3^3$ is the Jacobian of the change of variables. Actually, one can easily check that the denominator in Eqs.(\ref{eqf223}) and (\ref{eqh223}) has the simpler form
$E-k^2/(2m)-x+((p'_1)^2+(p'_2)^2+{\bf p'}_1\cdot {\bf p'}_2)/m$.
\subsubsection{$\Sigma^{(2,2,d+e+f)}$}
The calculation of the last three diagrams (d),(e) and (f) for $\Sigma^{(2,2)}$ is more involved. This is explained in Appendix \ref{appsigdef}. We also have
\begin{multline}
\Sigma_R^{(2,2,d+e+f)}(k,\omega)=z^2 F_{2,2,d+e+f}(k,\omega+\mu+i 0^+)+\\
z^3 H_{2,2,d+e+f}(k,\omega+\mu+i 0^+)
\end{multline}
The results are
\begin{multline}\label{eqsigdeffin1}
F_{2,2,d+e+f}(k,E)= {\mathcal P}\int\frac{d{\bf P}_t d{\bf p'}_1 d{\bf p'}_2}{(2\pi)^9}\int_0^{+\infty}\hspace{-5.mm}dx_1
\int_0^{+\infty}\ddx\\
\times\frac{e^{-\beta(x_1+\frac{({\bf P}_t-{\bf k})^2}{4 m})}}
{E-x-\frac{P_t^2}{6m}+x_1+\frac{({\bf P}_t-{\bf k})^2}{4m}
}
\frac{\rho_2(x_1)}{(x_1-\varepsilon'_1)(x_1-\varepsilon'_2)}\\
\left[
\rho_2(x-\frac{3}{4}\frac{(p'_1)^2}{m})\delta({\bf p}'_1-{\bf p}'_2)(2\pi)^3+\rho_3({\bf p}'_1,{\bf p}'_2;x)
\right]
\end{multline}
\begin{multline}\label{eqsigdeffin2}
H_{2,2,d+e+f}(k,E)= {\mathcal P}\int\frac{d{\bf P}_t d{\bf p'}_1 d{\bf p'}_2}{(2\pi)^9}\int_0^{+\infty}\hspace{-5.mm}dx_1
\int_0^{+\infty}\hspace{-5.mm}dx\\
\times\frac{e^{-\beta(x+\frac{{\bf P}_t^2}{6 m})}}
{\left[E-x-\frac{P_t^2}{6m}+x_1+\frac{({\bf P}_t-{\bf k})^2}{4m}
\right]}
\frac{\rho_2(x_1)}{(x_1-\varepsilon'_1)(x_1-\varepsilon'_2)}\\
\left[
\rho_2(x-\frac{3}{4}\frac{(p'_1)^2}{m})\delta({\bf p}'_1-{\bf p}'_2)(2\pi)^3+\rho_3({\bf p}'_1,{\bf p}'_2;x)
\right]
\end{multline}
where $\varepsilon'_{1,2}$ are functions of the integration wave vectors. We have $\varepsilon'_{1}=(p'_1)^2/m+(P_t)^2/(36 m)-{\bf p'_1}\cdot {\bf P}_t/(3m)+{\bf p'_1}\cdot {\bf k}/(m)-{\bf k}\cdot {\bf P}_t/(6m)+k^2/(4m)$ and an equivalent expression for $\varepsilon'_{2}$.

\section{Numerical results in the Unitary Limit}\label{numUL}
Our numerical calculations were performed in the unitary limit. This brings simplifications, since one can rescale the wave vectors entering in $t_3$, and there are only two variables in $t_3$, which significantly decreases the amount of computation. It is one of the reasons why we studied numerically the unitary limit. 
The imaginary part of the self-energy is calculated directly numerically, and the real part is obtained using Kramers-Kronig relations.
\subsection{High negative frequency, high wave vector behavior and Tan's Contact}
\begin{figure}[h]
\begin{center}
\includegraphics[width=.9\linewidth]{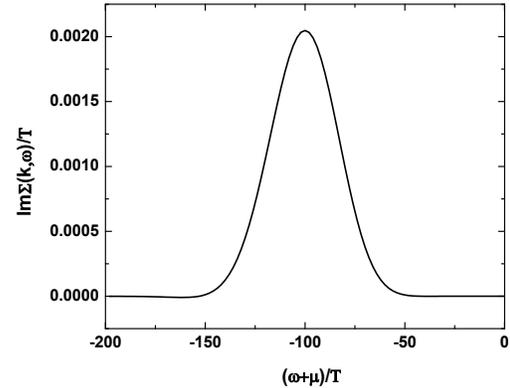}
\caption{The imaginary part of the self-energy at order $z^3$ (see text) for $k\Lambda_T=20\sqrt{\pi}$.}
\label{figimsigmac3}
\end{center}
\end{figure}

We first consider the large momentum $k$ and large negative $\omega$ limit of the self energy, where we expect to recover the physics of Tan's contact. In Fig.\ref{figimsigmac3} is shown the imaginary part of the self-energy at order $z^3$ (the imaginary part of the sum of $H_{2,1}$ and $H_{2,2,\cdots}$). In agreement with the lowest order calculation, we find that $F_{2,1}$ and $F_{2,2,\cdots}$ do not contribute. The numerical results show that the main contribution comes from $H_{2,2,a}$ and $H_{2,2,b}$, and can be well fitted by a Gaussian. Moreover, by integration on $\omega$, we get the occupation number $n_k$, from which we can find a value for the virial expansion coefficient $c_3$ (see Eq.(\ref{eqdefc3})) of Tan's contact at order $z^3$. We find a good agreement (within a few percent) with the value found in section \ref{seccontact3} at unitarity, using a virial expansion of the two-particle vertex $\Gamma$. We expect this feature to be valid at any order of the expansion in fugacity $z$.
\subsection{Spectral function $A(k,\omega)$ : effect of three-body correlations}
\begin{figure}[h]
\begin{center}
\subfigure[\label{imsig221UL}]{\includegraphics[width=0.8\linewidth]{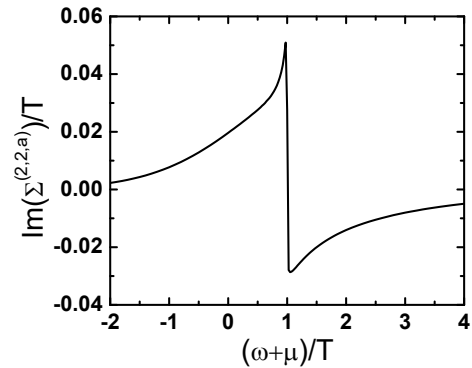}}\\
\subfigure[\label{imsig2UL}]{\includegraphics[width=0.8\linewidth]{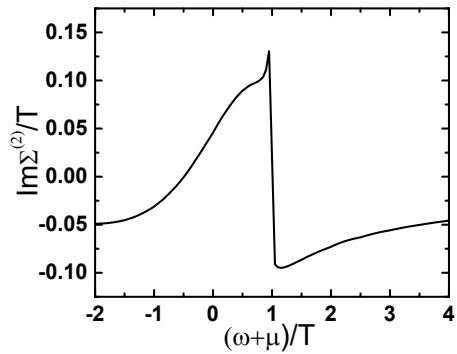}}
\caption{(a) Imaginary part of $\Sigma^{(2,2,a)}(k,\omega)$ for $k \Lambda_T=2\sqrt{\pi}$ and $z=0.3$. (b) Imaginary part of $\Sigma^{(2)}(k,\omega)$, the second order self-energy, for $k \Lambda_T=2\sqrt{\pi}$ and $z=0.3$.}
\label{figimsigma3}
\end{center}
\end{figure}

We come now to the main results of this work on the effect of three-body correlations on the spectral function. We have calculated numerically the retarded self-energies corresponding to diagrams of Figs.\ref{figSig2_1} and \ref{figsig23af}. 
In Fig.\ref{imsig221UL}, the imaginary part of $\Sigma^{(2,2,a)}$ ($F_{2,2,a}$ the contribution of order $z^2$) is plotted for $k \Lambda_T=2\sqrt{\pi}$ and $z=0.3$. Using the virial expansion of Appendix \ref{appfug} up to order $z^3$, we find $T/T_F=1.60$. We observe a sharp structure at the point $\omega+\mu=k^2/(2m)$. We have found a sharp variation of the form $\sqrt{|\omega+\mu-k^2/(2m)|}$. This is explained in Appendix \ref{struct221}. We found a more pronounced variation at $\omega+\mu>k^2/(2m)$ than at  $\omega+\mu<k^2/(2m)$ because the coefficient multiplying the square root is larger in the former case. Indeed, for the channel $l=0$ angular momentum channel, this ratio is about $8$. This singularity is associated to the $2$-body matrix $t_2(s)$ and as a result this singular behavior is observed in other terms of the second order self-energy. In Fig.\ref{imsig2UL}, we plot the imaginary part of the total self energy $\Sigma^{(2)}$ of order $z^2$ for the same wavevector and fugacity (including $O(z^2)$ contributions of diagrams of Figs.\ref{figsigma1}, \ref{figSig2_1} and \ref{figsig23af}). Here too, we can observe a sharp structure around $\omega+\mu=k^2/(2m)$. We also observe that the contribution to the imaginary part (see Fig.\ref{imsig2UL}) of the self-energy is positive, whereas the self-energy must have a negative imaginary part. Actually, this is consistent with Fig.\ref{figimsigmac3}, where the imaginary part of the self energy is also positive, consistent with the fact that $c_3$ is negative. However, this restricts the reasonable values of the fugacity since, for example, for $z=0.5$ and $k\Lambda_T=\sqrt{\pi}/5$, the spectral function can have some negative contributions.

\begin{figure}[h]
\begin{center}
\subfigure[\label{Dosk01UL}]{\includegraphics[width=0.8\linewidth]{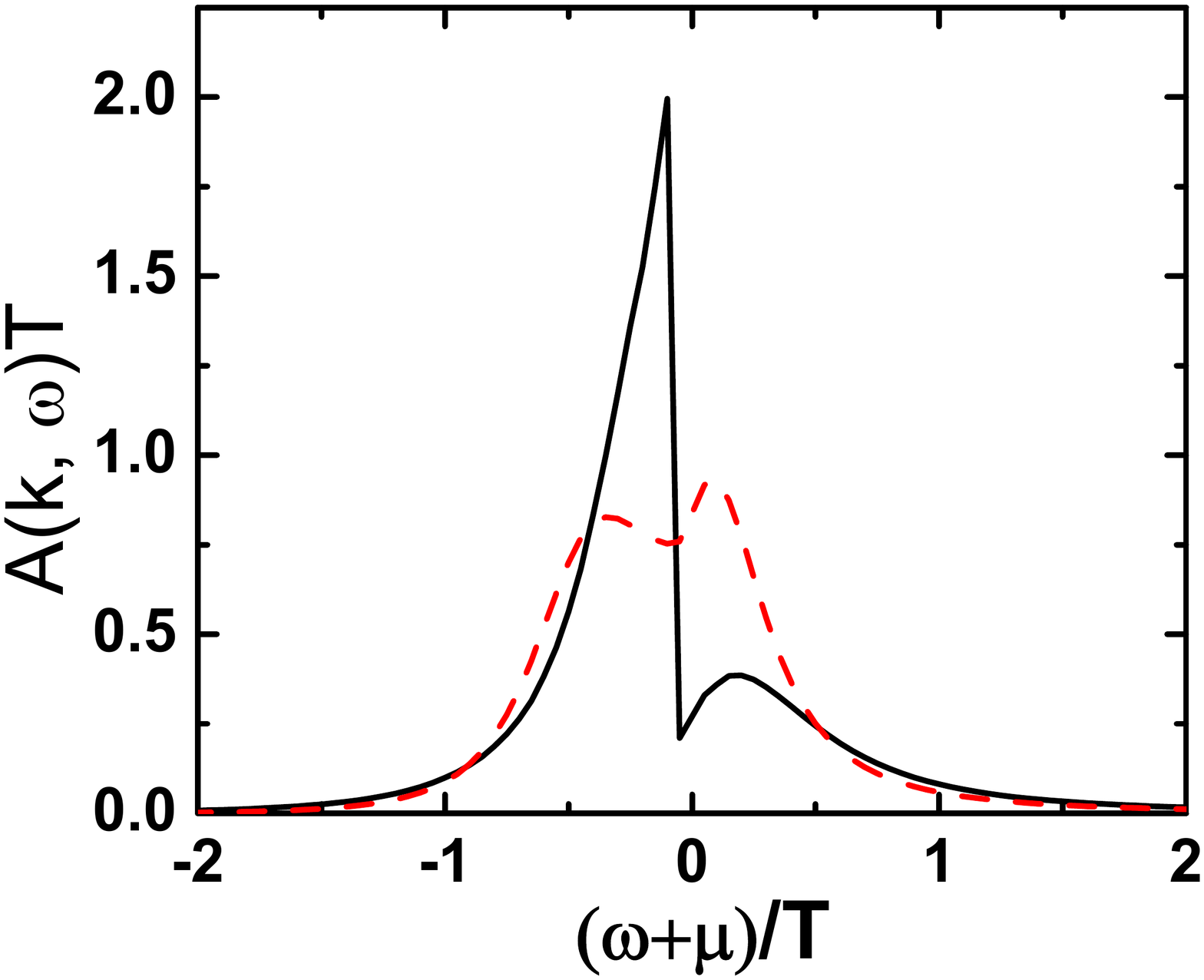}}\\
\subfigure[\label{Dosk1UL}]{\includegraphics[width=0.8\linewidth]{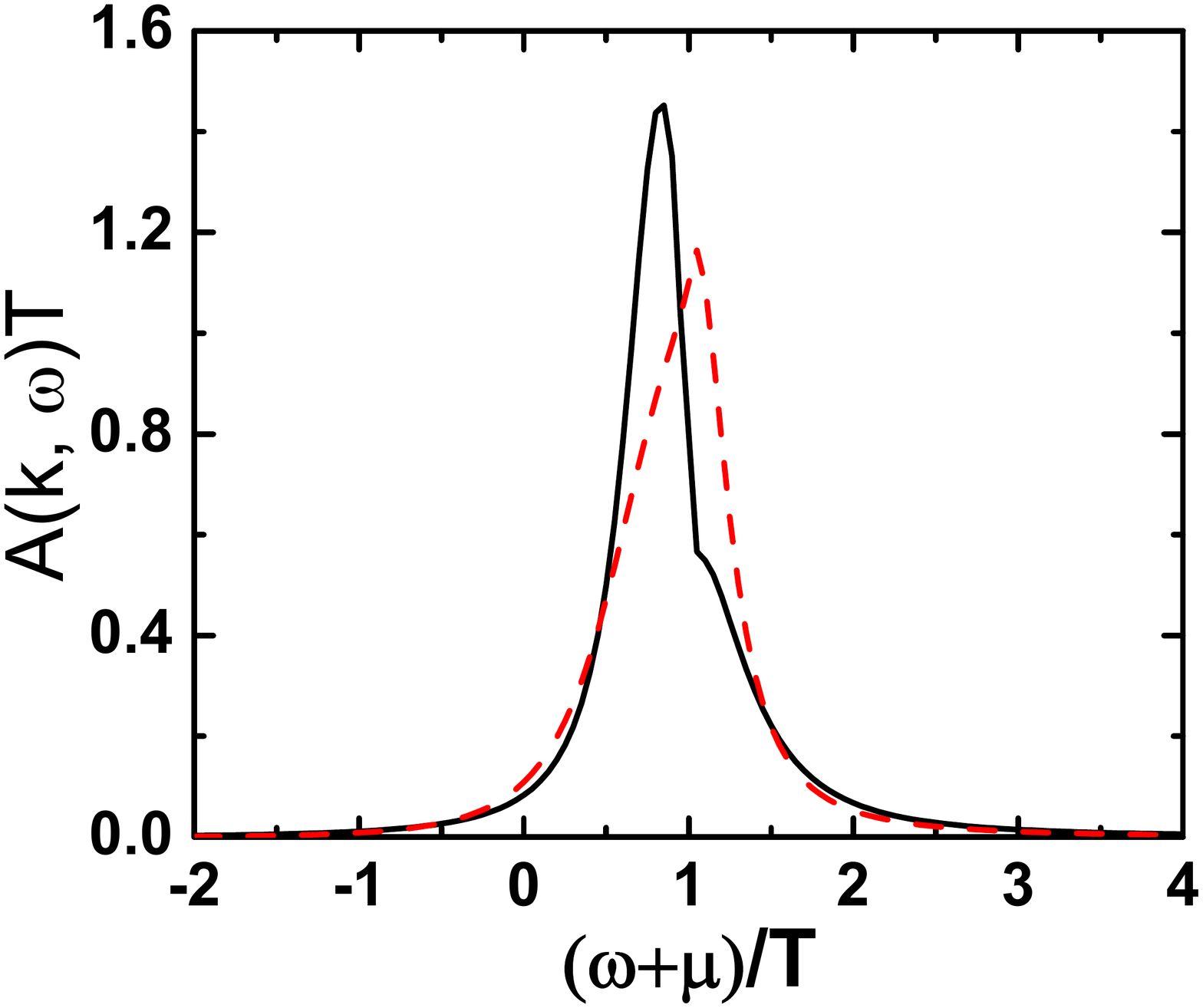}}\\
\subfigure[\label{Dosk2UL}]{\includegraphics[width=0.8\linewidth]{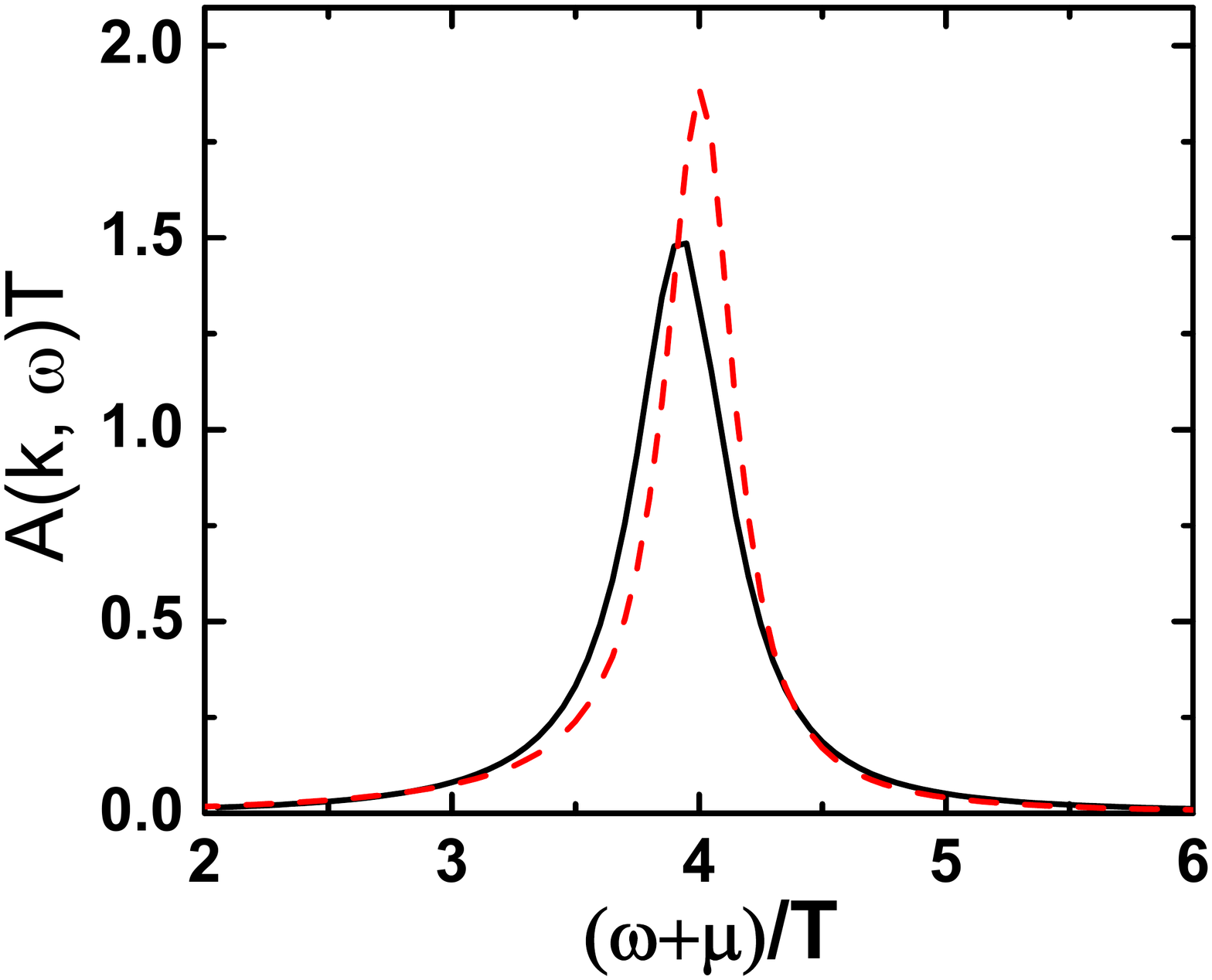}}
\caption{Spectral function at second order (blue solid line), compared to the lowest order ones (red dashed line).The fugacity is $z=0.3$ ($T/T_F=1.60$). (a) $k\Lambda_T=\sqrt{\pi}/5$, (b) $k\Lambda_T=2\sqrt{\pi}$, (c) $k\Lambda_T=4 \sqrt{\pi}$}
\label{figAk12}
\end{center}
\end{figure}

The second order spectral function (including all $O(z^2)$ terms in the self-energy) is plotted in Fig.\ref{figAk12} for $z=0.3$.
Given the features of the imaginary part of the second order self-energy, we find a very different behavior compared to the lowest order calculation \cite{pieristrinatipg}.  Similar to lowest order, there exists a two-peak structure at small momentum, which gradually evolves into one peak at large momentum. However, the two peaks are very asymmetric and the left peak dominates. This is shown in Fig.\ref{Dosk01UL}, for $z=0.3$ and $k\Lambda_T=\sqrt{\pi}/5$. We conclude that $3$-body correlations can be quantitatively important for the spectral function, already at a high temperature corresponding to $z=0.3$. Naturally, we expect this will become more pronounced when the temperature decreases, and the fugacity increases. Our calculations imply that it is not sufficient to include only $2$-body correlations in order to obtain the correct spectral function, in particular close to the critical temperature.

Concerning the issue of convergence of our results with respect to higher powers of the fugacity $z$, the only thing we can safely say is that at this value of $z$ ($=0.3$), the virial expansion of the equation of state at third order reproduces accurately the experimental results of Ref.\cite{natureeoslkbli6}. Hence, we may expect that the expansion has also converged for the spectral function.

\section{Conclusion}\label{conclusion}
In this work, we have extended a high temperature, virial-like, expansion to a dynamical quantity such as the self-energy or the spectral function. Even though the method can be applied to bosons as well as fermions, we have considered in detail the BEC-BCS crossover for spin $1/2$ fermions in $3D$. We have calculated the self-energy on the real axis up to order $z^2$ ($z$ the fugacity), including three-body correlations, for $a^{-1}\leq 0$ (Unitary-BCS side). We have performed numerical calculations in order to evaluate the spectral function in the unitary limit, including three-body correlations. We find a quantitative importance of three-body correlations even in this high temperature regime. We have also calculated the third order ($O(z^3)$) virial expansion of Tan's contact, in the whole BEC-BCS crossover.

The problem of the mixture of fermions of different masses is a natural extension of our work. Experimentally and theoretically, this has been studied in \cite{grimmpetrovlevinsen} with a mixture of $^6 Li$ and $^{40}K$ atoms. There, we expect $3$-body correlations to become particularly important since, at least on BEC side, one expects \cite{mk3bl1} a three-body bound state in $l=1$ channel for mass ratio around $8$.

\acknowledgements
We thank N. Dupuis for  a critical reading of the article, K. van Houcke and F. Werner for fruitful discussions and comparison to diagrammatic MC calculations.
 \appendix
 \section{Finding the fugacity for given $T/T_F$ and $1/(k_F a)$}\label{appfug}
The goal of this section is to explain how one can find the two dimensionless parameters $z$ (the fugacity)
and $\Lambda_T/a$ from the two other dimensionless parameters $T/T_F$ and $1/(k_F a)$.

The Fermi momentum $k_F$ is defined such that the total density $n=(k_F^3)/(3\pi^2)$ and the Fermi temperature is defined by $T_F=k_F^2/(2m)$. $\Lambda_T=\sqrt{2\pi/(m\,T)}$ is the thermal wavelength. The virial  expansion up to order $z^2$ states that
$n\Lambda_T^3=2\left[z+2\, b_2(\Lambda_T/a)z^2\right]$. From this, we get the two equations
\bea
\frac{\Lambda_T}{a}&=&2\sqrt{\pi}\frac{1}{k_F\,a}\left(\frac{T}{T_F}\right)^{-1/2}\label{eqfug1}\\
z+2\,b_2(\frac{\Lambda_T}{a})z^2&=&\frac{4}{3\sqrt{\pi}}\left(\frac{T}{T_F}\right)^{-3/2}\label{eqfug2}
\eea
For given values of $1/(k_F\,a)$ and $T/T_F$, we find $\frac{\Lambda_T}{a}$ from Eq.(\ref{eqfug1}). This gives the virial cumulant $b_2(\Lambda_T/a)$ (see for  instance \cite{pravirial} for its expression \cite{erratumpra}) in Eq.(\ref{eqfug2}), which for given $T/T_F$ becomes a (second order) equation for the fugacity $z$. In this way, if $T/T_F=1$, we find $z=0.17$ for $1/(k_F\,a)=1$ (BEC side),
$z=0.5$ for $1/(k_F\,a)=0$ (unitary limit) and $z=0.7$ for $1/(k_F\,a)=-1$ (BCS side).

 \section{BCS limit}\label{appendixBCSlimit}
The BCS limit is a weak coupling limit, therefore we expect that the real part of the self-energy will be given by a mean-field like term $g n_{\downarrow}$, where $g=4\pi a/m$ is the coupling constant and at lowest order $n_{\downarrow}=\,z(m T/(2\pi))^{3/2}$. This yields an energy shift $\sqrt{(2 m)/\pi}a\, z\,T^{3/2}$.  Notice that this shift is negative, as can be seen in Fig.\ref{spectralfunction-1}. We can recover this results using the explicit expression of the self-energy $\Sigma(k,\omega)$ in the limit $a^{-1}\to-\infty$.  The width of the peak is obtained using a Fermi Golden Rule approach.
The initial state is obtained by the creation of a spin $\uparrow$ fermion in a gas of $2 N$ classical fermions ($N$ spin $\uparrow$, $N$ spin $\downarrow$). Let us denote the momenta of the spin $\downarrow$ fermions ${\bf p}_1,\cdots,{\bf p}_N$. The created fermion scatters on a spin $\downarrow$ fermion and the final states are obtained by the created fermion with a momentum ${\bf k+q}$ while the scattered spin $\downarrow$ fermion, for instance the one of initial momentum ${\bf p}_N$ has now a momentum ${\bf p}_N-{\bf q}$, by momentum conservation. All the other particles keep their momenta.
The difference in energy in this process is $({\bf k}+{\bf q})^2/(2m)+({\bf p}_N-{\bf q})^2/(2m)-(k^2/(2m)+p_N^2/(2m))=q^2/m+({\bf k}-{\bf p}_N)\cdot {\bf q}/m$.
The inverse of the lifetime of the initial state is therefore
\bea
\frac{1}{\tau_i}&=&2\pi g^2 \dq \delta\left(q^2/m+({\bf k}-{\bf p}_N)\cdot {\bf q}/m)\right)\nonumber\\
&=&\frac{1}{4\pi} g^2 m |{\bf k}-{\bf p}_N|\label{eqtaui}
\eea 
The next step is to average on the fermions state, using a Boltzmann distribution. The calculation is similar to the one of Appendix \ref{A-BEC}, and we just give the result
\bea
\frac{1}{\tau_k}&=&\frac{1}{\pi}\frac{ z\,a^2}{\beta\,k}\int_0^{+\infty} dp p^2\left[e^{-\beta\frac{(k-p)^2}{2m}}-e^{-\beta\frac{(k+p)^2}{2m}}\right]\label{eqtauk}
\eea
This result can actually be recovered from the expression of the imaginary part of $\Sigma(k,\omega)$ at lowest order in $z$ in the limit $a^{-1}\to -\infty$, for $\omega+\mu=k^2/(2m)$. Indeed at lowest order in the fugacity $z$, the imaginary part of the self energy is the imaginary part of $F_1(k,\omega+\mu+i 0^+)$. In the BCS limit $a^{-1}\to -\infty$, and we can replace $\rho_2(x)$ in Eq.(\ref{eqF_1}) by $4/m^{1/2} a^2\sqrt{x}$. In a perturbative approach, we can also replace $\omega+\mu$ by $k^2/(2m)$ and we find
for $-\Im(\Sigma(k,k^2/(2m)+i0^+))$ at lowest order in $z$ and $a$
$$
\pi z\hspace{-2.mm}\dP e^{-\beta\frac{({\bf P-k})^2}{2m}}\hspace{-2.mm}\int_0^{+\infty}\ddx\frac{4}{m^{1/2}}a^2\sqrt{x}
\delta(\frac{k^2}{2m}-x+\frac{({\bf P-k})^2}{2m}-\frac{P^2}{4m}
)
$$
The integral on $x$ can be performed, and by comparison with Eq.(\ref{eqtauk}), we find the expected result (after the change of variable ${\bf P}'=2 {\bf k}-{\bf P}$) $-2\Im(\Sigma(k,k^2/(2m)+i0^+))=1/\tau_k$.

When we compare our numerical results with the perturbative results obtained in the BCS limit, we see that we need to go quite far in the BCS regime in order to get a good quantitative agreement.

 \section{BEC limit}\label{appendixBEClimit}
 \subsection{Occupied spectral function of a gas of classical dimers}\label{A-BEC}
 We recall the definition of the occupied spectral function $A_{-}(k,\omega)$
 \bea
 A_{-}(k,\omega)&=&\frac{1}{Z_{\mu}}\sum_{n,m}e^{-\beta E_n}|\langle m|c_{{\bf k},\uparrow}|n\rangle|^2
 \delta(\omega+E_m-E_n)\nonumber\\
 \eea
 where the sum extends to all the eigenstates $|n\rangle$ and $|m\rangle$ of $H'=H-\mu N$ where $H$ is the Hamiltonian of the system and $N$ the particle number operator. $Z_{\mu}$ is the grand canonical partition function.
 In the BEC limit, the eigenstates $|n\rangle$ are given by $\uparrow\downarrow$ dimers in states of momenta ${\bf P}_1,\cdots,{\bf P}_M$ (states with an odd number of atoms have much higher energy since they don't benefit of  the binding energy). For a given dimer number $M$, the action of the annihilation operator $c_{{\bf k},\uparrow}$ couples to states consisting of a single $\downarrow$ particle of momentum ${\bf P_i-k}$ ($i=1,\cdots M$) and dimers in states ${\bf P}_1,\cdots,{\bf P}_{i-1},{\bf P}_{i+1},\cdots{\bf P}_M$.
 Therefore we have $E_m-E_n=\mu+({\bf P_i-k})^2/(2m)+E_b-P_i^2/(4m)$. The indistinguishability of the dimers makes that there is a factor $M$ (the number of possible values of $i$) in the summation. We choose $P_M$ as the momentum variable of the dissociated dimer. The matrix element in the dilute limit is (we basically ignore the effect of the $M-1$ dimers)
 $\langle m|c_{{\bf k},\uparrow}|n\rangle\simeq\langle {\bf P_M-k}\downarrow|c_{{\bf k},\uparrow}|D:{\bf P}_M\rangle$, where the dimer state of momentum ${\bf P}_M$ is
 $|D:{\bf P}_M\rangle=\sum_{{\bf q}}\varphi_{{\bf q}}c^{\dagger}_{{\bf P}_M/2+{\bf q}\uparrow}c^{\dagger}_{{\bf P}_M/2-{\bf q}\downarrow}|0\rangle$ ($|0\rangle$ is the vacuum state).
 $\varphi_{{\bf q}}=(8\pi/a)^{1/2}1/(q^2+a^{-2})$ is the dimer wave function. We find for the matrix element $\langle m|c_{{\bf k},\uparrow}|n\rangle\simeq\varphi_{{\bf k-P}_M/2}$. The summation on the $M-1$ dimer variables give the partition function of $M-1$ classical particles of mass $2m$. It factorizes from the rest and is $\sum_{M>1} M/M!(\sum_{{\bf P}}e^{-\beta(P^2/(4m)-E_b-2\mu)})^{M-1}=Z_{\mu}$ which simplifies with the $Z_{\mu}$ at the denominator. We find
 \begin{multline}
 A_{-}(k,\omega)=\dP e^{-\beta(\frac{P^2}{4m}-E_b-2\mu)}\\ \frac{8\pi}{a}\frac{\delta(\omega+\mu+\frac{({\bf P-k})^2}{2m}+E_b-\frac{P^2}{4m})}{\left(a^{-2}+({\bf k-P}/2)^2\right)^2}
 \nonumber\\
 \end{multline}
 This can be written, due to the Dirac function
 \begin{multline}
 A_{-}(k,\omega)=z^2\frac{8\pi}{m^2 a}e^{\beta E_b}\frac{1}{\left(\omega+\mu-\frac{k^2}{2m}\right)^2}\\
 \dP e^{-\beta\frac{{\bf P}^2}{4m}}\delta(\omega+\mu+\frac{({\bf P-k})^2}{2m}+E_b-\frac{P^2}{4m})\label{eqAmbec}
 \end{multline}
From this expression, we easily find the threshold $$\omega+\mu\le -E_b+k^2/(2m)$$ The integral in Eq.(\ref{eqAmbec}) can be performed analytically, and we find
\bea
 A_{-}(k,\omega)&=&z^2\frac{T}{\pi a\, k}e^{\beta E_b}
 \frac{
 e^{
 -\frac{\beta P_{-}^2}{4 m}
 }
 -e^{-\frac{\beta P_{+}^2}{4 m}}
 }
 {
 (\omega+\mu-\frac{k^2}{2m})^2}
\eea
if $\omega+\mu\le -E_b+k^2/(2m)$ and $0$ else. We have $P_{-}=2(k-\sqrt{m |\omega+\mu+E_b-k^2/(2m)|})$ and  $P_{+}=2(k+\sqrt{m |\omega+\mu+E_b-k^2/(2m)|})$.
\subsection{Particle peak}
At lowest order, the situation is similar to the BCS limit. One finds a positive mean field like shift of the peak located around $\omega+\mu=k^2/(2m)$, and a broadening given by the Fermi Golden rule like calculation Eq.(\ref{eqtauk}).
 \section{Limit of $\Im[\Sigma^{(1)}_R(k,\omega)]$, $\omega\to -\infty$, $k\to \infty$}\label{appendix high k}
 We calculate the dominant contribution of the imaginary part of $-1/\pi H_1(k,\omega+\mu+i\,0^+)$ in the limit of high $k$ and large non positive frequency $\omega+\mu$.
 Due to the Boltzmann weights $e^{-\beta P^2/(4m)}$ and $e^{-\beta x}$, in this limit, we find that $P\sim (m/\beta)^{1/2}$ and $x\leq (m/\beta)^{1/2}$. This means that in this limit
 $P^2/4m\ll |\omega+\mu|, k^2/m$ and $x\ll  |\omega+\mu|, k^2/m$. Hence we find
\begin{multline}
 -\frac{1}{\pi}\Im[H_1(k,\omega+\mu+i\,0^+)]\simeq\dP e^{-\beta\frac{P^2}{4m}}\times\\
 C'\delta(\omega+\mu+\frac{k^2}{2 m}-\frac{{\bf P}\cdot{\bf k}}{m})\\
 \end{multline}
 where the constant $C'$ is given by
 \bea
 C'&=&\int_0^{+\infty}\ddx e^{-\beta x}\rho_2(x)+\Theta(a^{-1})e^{\beta E_b}Z_m
 \eea
 The reason why we do not neglect $\frac{{\bf P}\cdot{\bf k}}{m}$ is that $k$ is large in this limit and this scalar product can be comparable to $\omega+\mu+k^2/(2m)$ in the relevant case where $\omega+\mu\simeq -k^2/(2m)$. The integral on ${\bf P}$  can be performed exactly and we find
 \bea
  -\frac{1}{\pi}\Im[H_1(k,\omega+\mu+i\,0^+)]&\simeq&\frac{1}{2 \pi^2}\frac{m^2}{\beta k} C' 
  \exp(-
  \frac{(\omega+\mu+\frac{k^2}{2m})^2}{4 T\frac{k^2}{m}}
  )
  \nonumber\\
 \eea
 which is a Gaussian centered around $\omega+\mu=-k^2/(2m)$ and of width of order $(T/m)^{1/2} k$.

\section{Unfolding trick}\label{unfold}
\begin{figure}[h]
\subfigure[\label{Figunfold1}]{\includegraphics[width=0.65\linewidth]{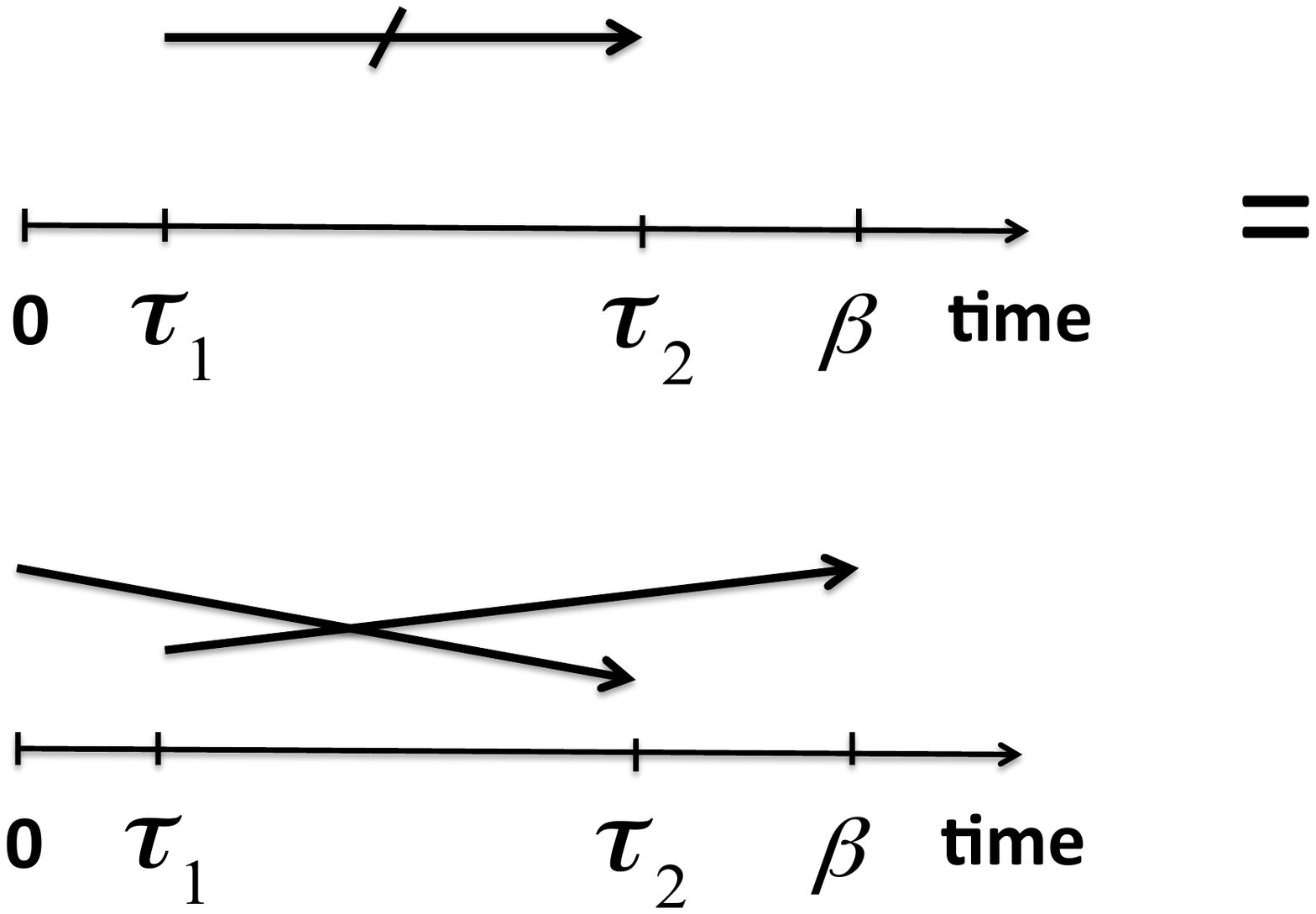}}\\
\subfigure[\label{Figunfold2}]{\includegraphics[width=0.65\linewidth]{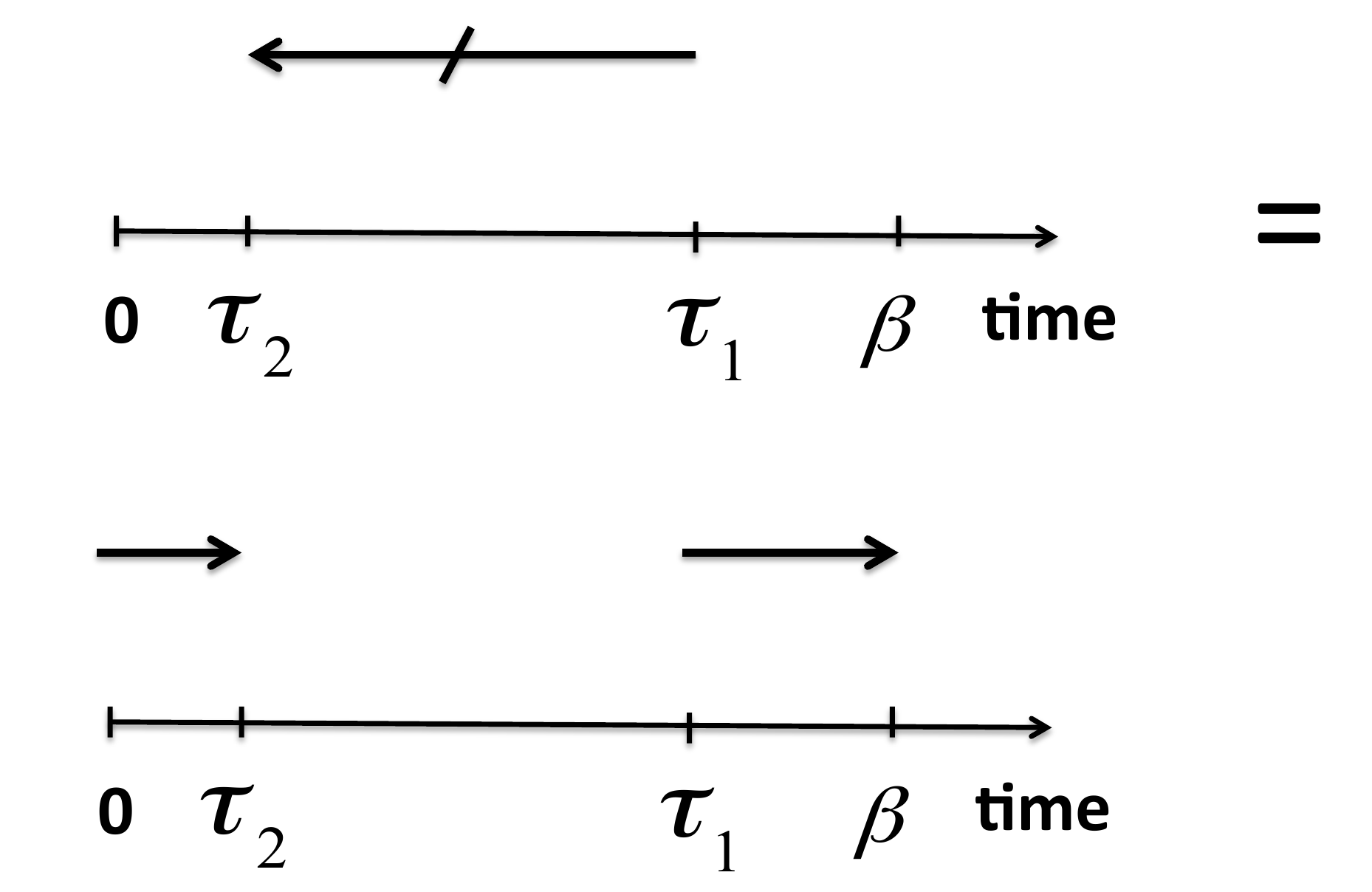}}
\caption{Unfolding trick : the slashed line $G^{(0,1)}(\tau_2-\tau_1)$ can be drawn as the product of two unslashed $G^{(0,0)}$ lines. (a): $\tau_2>\tau_1$, (b): $\tau_2<\tau_1$.
}
\end{figure}

We show here how a slashed propagator $G^{(0,1)}({\bf p},\tau)$ can be reinterpreted  in terms of Feynman diagrams.
We consider a general slashed fermionic line between two times $\tau_1$ and $\tau_2$ with $0\leq \tau_1<\beta$ and $0\leq \tau_2<\beta$.
We have $G^{(0,1)}({\bf p},\tau_2-\tau_1)=e^{-\beta\varepsilon_{p}}e^{-(\tau_2-\tau_1)\varepsilon_{p}}=G^{(0,0)}({\bf p},\beta-\tau_1)G^{(0,0)}({\bf p},\tau_2)$. Therefore, a slashed line connecting the time $\tau_1$ to the time $\tau_2$ can be replaced by two unslashed $G^{(0,0)}$ lines. The first connects the times $\tau_1$ and $\beta$ while the second $G^{(0,0)}$ line connects the times $0$  and $\tau_2$. This is shown in Fig.\ref{Figunfold1} and \ref{Figunfold2}.

\section{Calculation of $\Gamma^{(3)}({\bf P},\beta^-)$}\label{appendixGamma2T_3}
The diagrams of Figs.\ref{figGamma2_T3a},\ref{figGamma2_T3b} can be drawn using the "unfolding" trick (see Appendix \ref{unfold}) as shown in Figs.\ref{figGamma2_a_unfold},\ref{figGamma2_b_unfold}.
\begin{figure}[h]
\subfigure[\label{figGamma2_a_unfold}]{\includegraphics[width=0.65\linewidth]{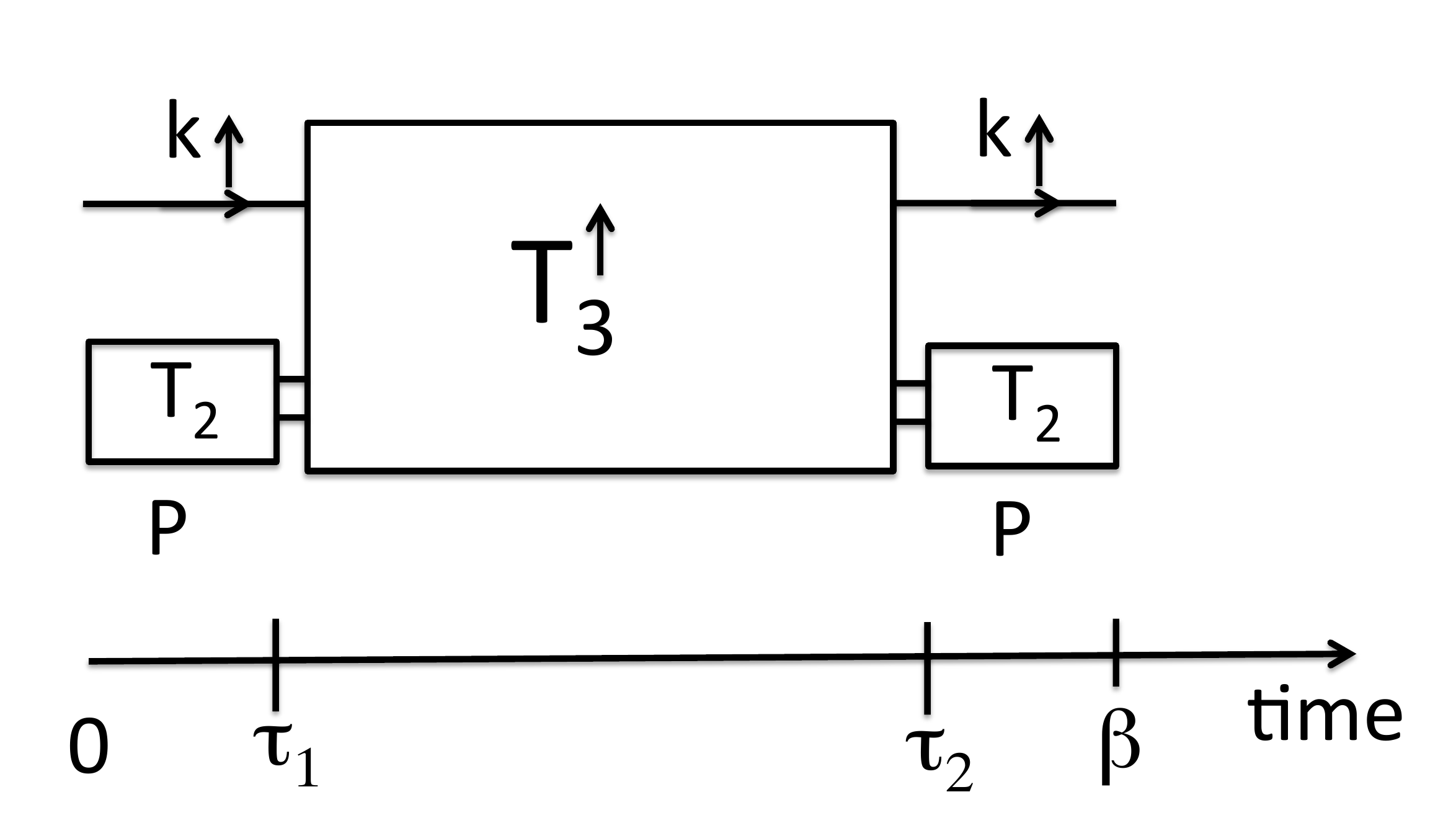}}\\
\subfigure[\label{figGamma2_b_unfold}]{\includegraphics[width=0.65\linewidth]{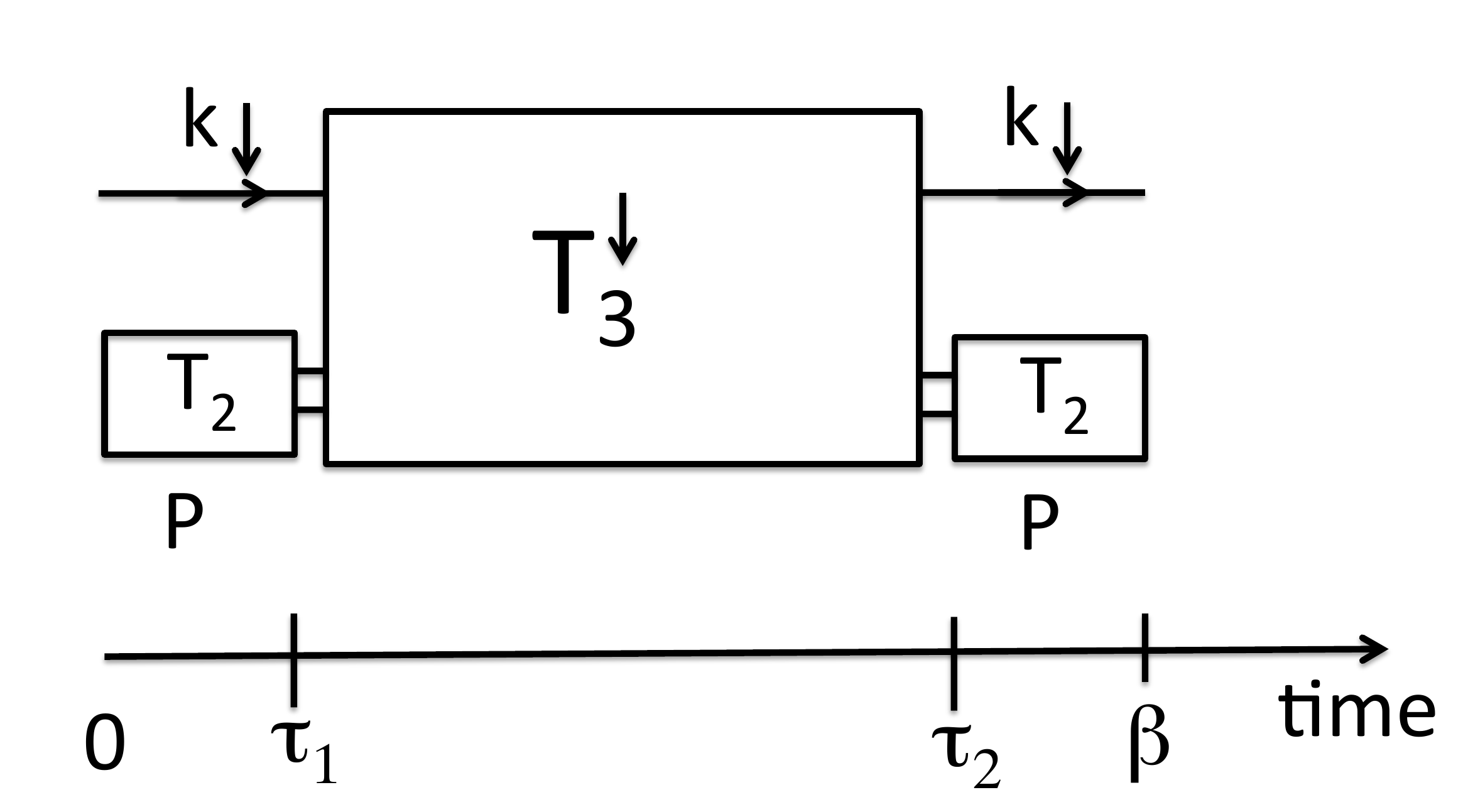}}
\caption{The two diagrams of Figs.\ref{figGamma2_T3a},\ref{figGamma2_T3b}, drawn in the unfolded way.
}
\end{figure}
The analytical expression in the imaginary time domain involves the intermediate times $\tau_1$ and $\tau_2$ as shown in Fig.\ref{figGamma2_a_unfold}. $\tau_1$ is the time of the last interaction in the incoming $T_2$, while $\tau_2$ is the  time of  the first interaction in the outgoing $T_2$. The whole diagram, {\it e.g.} Fig.\ref{figGamma2_a_unfold}  is a convolution product which can be written schematically, without wave vector indices, 
$$z^3\int_{\mathcal{D}}\hspace{-1.mm}dt_1\, dt_2 e^{-\varepsilon_{\uparrow}t_1}T_2(t_1)T_3(t_2)e^{-\varepsilon_{\uparrow}(\beta-t_1-t_2)}T_2(\beta-t_1-t_2)$$
where $t_1=\tau_1$ and $t_2=\tau_2-\tau_1$ and the integration domain
$\mathcal{D}=\{(t_1,t_2)\,:\,t_1\geq 0,t_2\geq 0, \beta-t_1-t_2\geq 0\}
$.
 This convolution product can be written using Laplace transforms \cite{pravirial}. For the diagram of Fig.\ref{figGamma2_a_unfold} we find
\bea
z^3&&\dk\ds e^{-\beta s}[T_2({\bf P},s-\varepsilon_k)]^2\nonumber\\
&&\times T_{3,\uparrow}[({\bf k},\varepsilon_k),({\bf k},\varepsilon_k);({\bf k+P},s)]\nonumber
\eea
In the Bromwich contour ${\mathcal C}_{\gamma}$, $\gamma$ is such that the integrand is analytical for $\Re(s)<\gamma$. The summation on ${\bf k}$ is done by going to the center of mass reference frame of the $3$-body problem as is explained in \cite{pravirial}.  $T_{3,\uparrow}$ is the $3$-particle $T$-matrix as defined in terms of Feynman diagrams in \cite{pra4par}.
We find for $\Gamma^{(3)}({\bf P},\beta^-)$
\begin{multline}
\Gamma^{(3)}({\bf P},\beta^-)=-z^3\int_{0}^{+\infty}\hspace{-5.mm}dp_1\frac{9\,m\, p_1}{8\pi^2\beta\,P}\\
\times\left[e^{-\frac{3}{8}\frac{\beta(p_1+P)^2}{m}} -(p_1\to -p_1) \right]\\
\times\left[\sum_{l\geq 0}\ds e^{-\beta s}\left(t_2(s-\frac{3}{4}\frac{p_1^2}{m})\right)^2\,t_{3,l}^{\uparrow}(p_1,p_1;s)\right]
\end{multline}
\section{Calculation of $\Sigma^{(2,2,d+e+f)}$}\label{appsigdef}
\begin{figure}[h]
\subfigure[\label{figappsig2d}]{\includegraphics[width=1.1\linewidth]{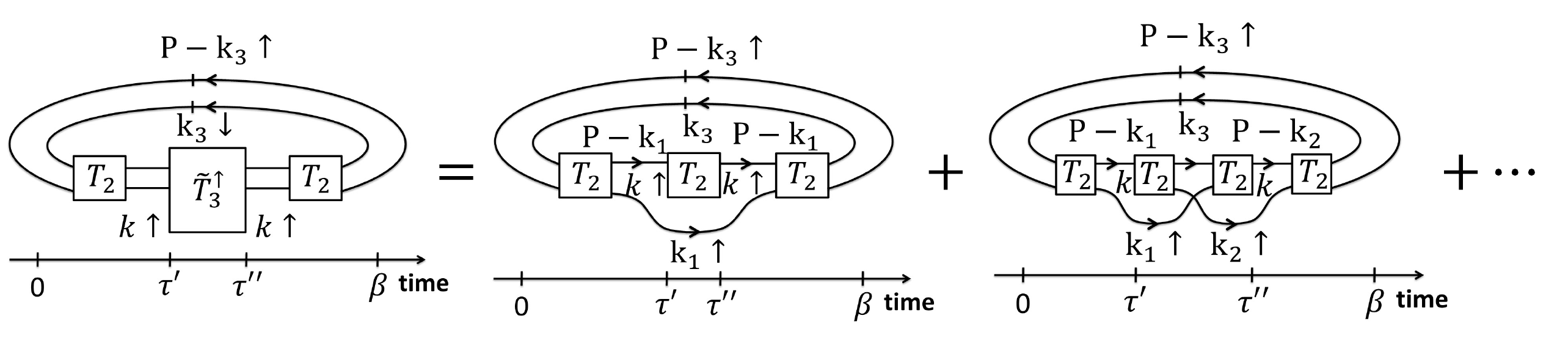}}\\
\subfigure[\label{figappsig2e}]{\includegraphics[width=1.1\linewidth]{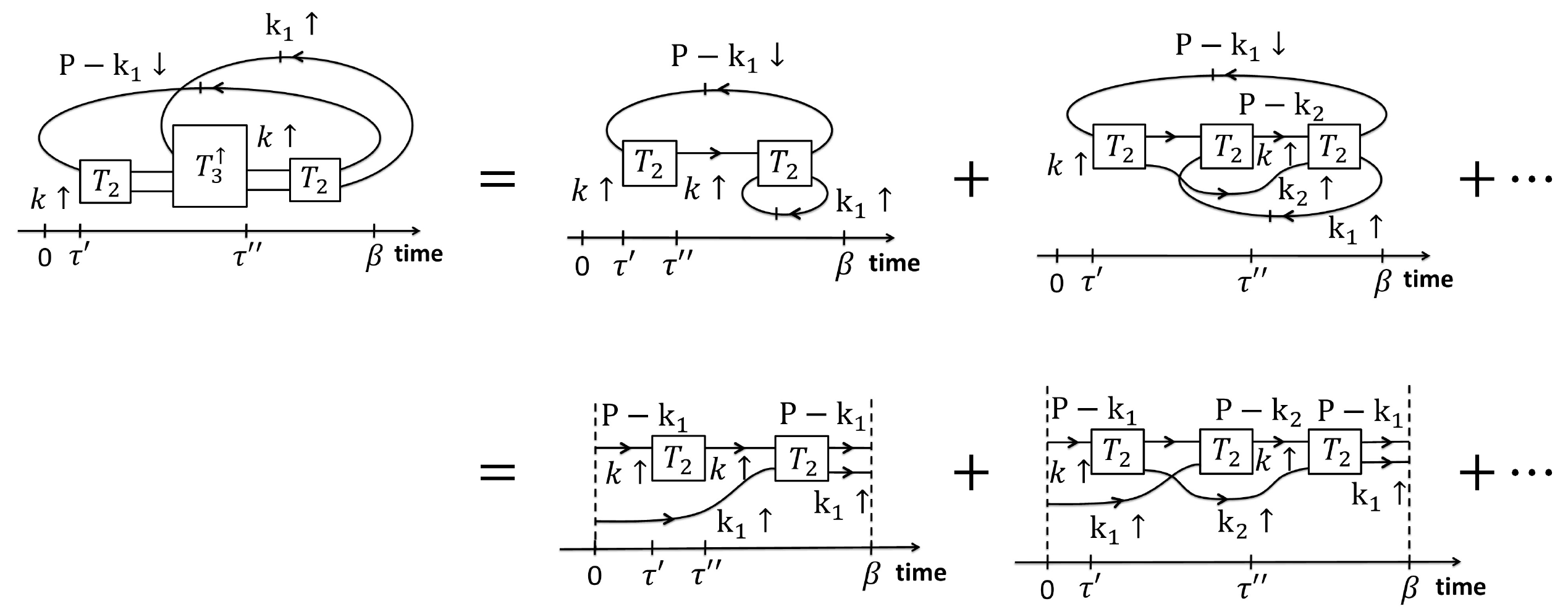}}\\
\subfigure[\label{figappsig2f}]{\includegraphics[width=1.1\linewidth]{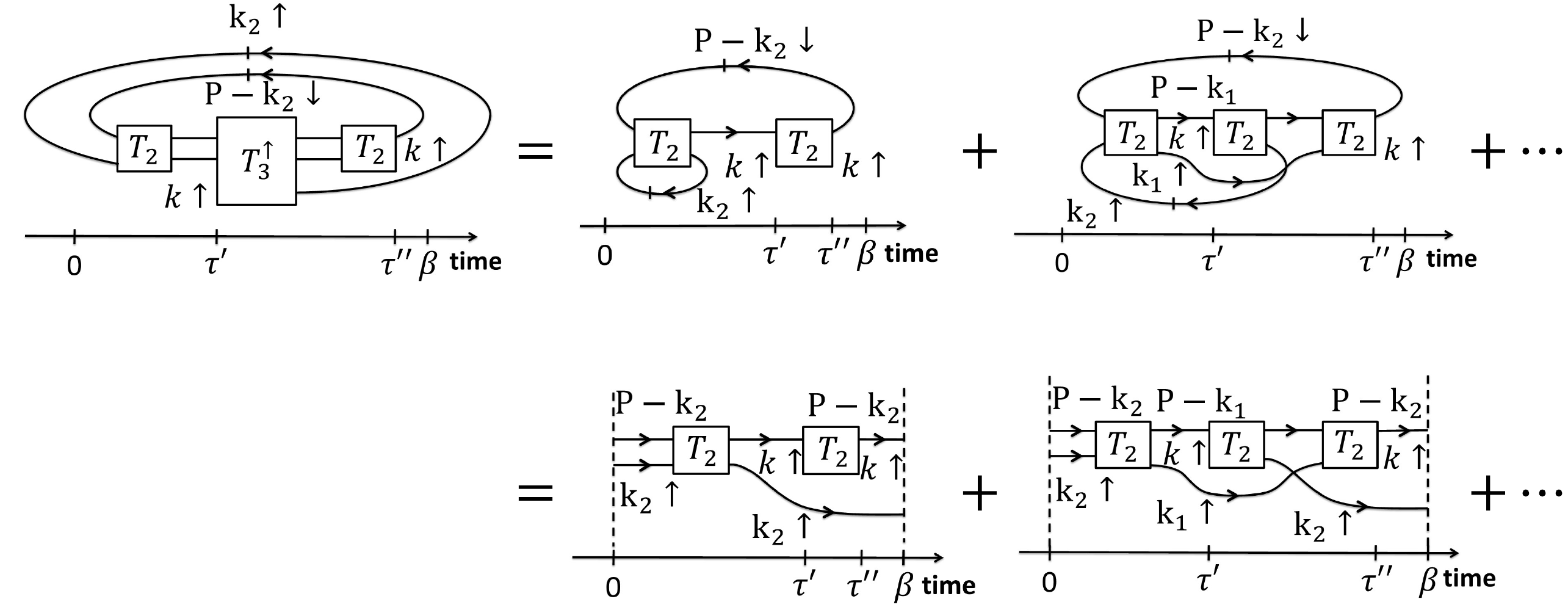}}\\
\subfigure[\label{figappsig2sum}]{\includegraphics[width=1.1\linewidth]{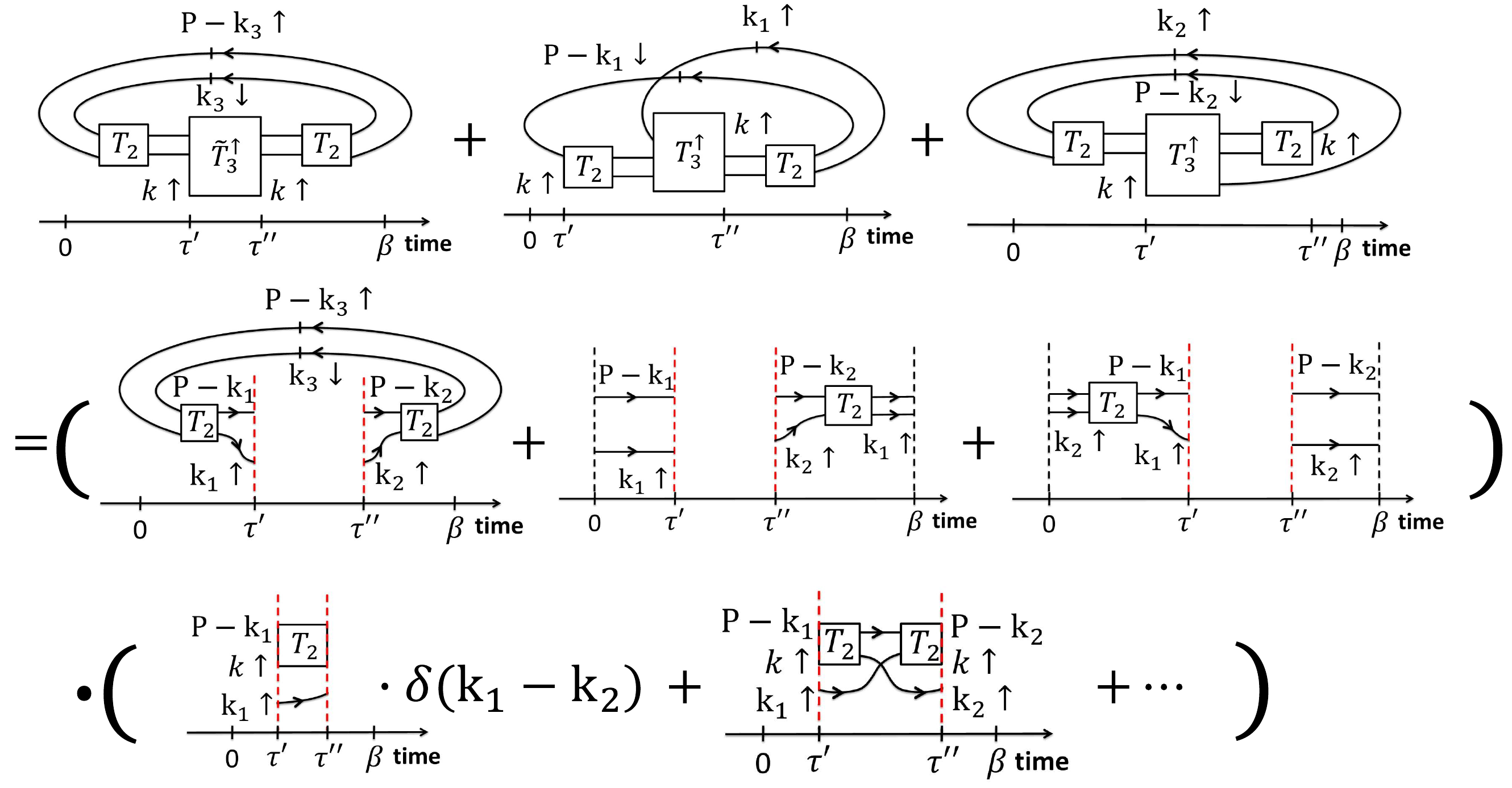}}
\caption{Diagrams of Figs.\ref{figsig23d},\ref{figsig23e} and \ref{figsig23f} written in the unfolded way. (a) is for $\Sigma^{2,2,d}$, 
(b) is for $\Sigma^{2,2,e}$ and (c) is for $\Sigma^{2,2,f}$.
(d) shows how they can be factorized.}
\end{figure}

The summation of the three diagrams of Figs.\ref{figsig23d},\ref{figsig23e} and \ref{figsig23f} is shown in Fig.\ref{figappsig2d} to Fig.\ref{figappsig2sum}.
There we see that the sum of the three diagrams can be factorized. 
The first term in the first parenthesis of Fig.\ref{figappsig2sum}  can be simplified. Indeed it contains the product of two $T_2$ and two $G^{(0,1)}$. If we denote by $\tau_2$ the time at the exit of the $T_2$ on the left and $\tau_1$ the time at the entrance of the $T_2$ on the right, we have the contribution
\begin{multline}
I({\bf P};\tau_2,\tau_1)=\dk\int_{{\mathcal D}} dx_1 dx_2 T_2({\bf P},x_1-\tau_1)\\
\times G^{(0,1)}({\bf k},x_2-x_1)G^{(0,1)}({\bf P-k},x_2-x_1)
T_2({\bf P},\tau_2-x_2)
\end{multline}
where ${\mathcal D}=\{(x_1,x_2),0<x_2<\tau_2<\tau_1<x_1<\beta\}$.
It is shown in Appendix \ref{eqT2T2p} that 
\begin{equation}
I({\bf P};\tau_2,\tau_1)=-T_2({\bf P},\beta+\tau_2-\tau_1)\label{eqT2T2}
\end{equation}

The sum of the three terms in the first parenthesis of Fig.\ref{figappsig2sum} is equal to (for given wave vectors ${\bf k}_1$ and ${\bf k}_2$)
\begin{multline}
-\int_{{\mathcal D}_1}dt_1dt_2 T_2({\bf P},\beta-\tau-t_1-t_2)e^{-(\varepsilon_1 t_1+\varepsilon_2 t_2)}\\
-\int_{{\mathcal D}_2}dt_1dt_2 T_2({\bf P},t_2)e^{-(\varepsilon_1(\beta-\tau-t_1- t_2)+\varepsilon_2 t_1)}\\
-\int_{{\mathcal D}_3}dt_1dt_2 T_2({\bf P},t_2)e^{-(\varepsilon_1 t_1+\varepsilon_2 (\beta-\tau-t_1-t_2))}
\end{multline}
where $\varepsilon_{1,2}=(k_{1,2}^2+({\bf P}-{\bf k}_{1,2})^2)/(2m)$, $\tau=\tau"-\tau'$ and $t_{1,2}$ denotes time differences.
In the second term, we make the change of variables $t'_1=\beta-\tau-t_1-t_2$ and $t'_2=t_1$. We get a new integration domain ${\mathcal D}'_2$. In the third term, we make the change of variables $t'_2=\beta-\tau-t_1-t_2$ and $t'_1=t_1$, and we also get a new integration domain ${\mathcal D}'_3$. By doing so, we find the same integrand for the three terms, and the integral must be down on the union of the three domains ${\mathcal D}={\mathcal D}_1\cup {\mathcal D}'_2\cup {\mathcal D}'_3$. We easily find
${\mathcal D}=\{(t_1,t_2),t_1\geq 0, t_2 \geq 0, \beta-\tau-t_1-t_2\geq 0  \}$. Therefore the sum is a convolution product and is equal to
\begin{equation}
\ds-t_2(s-\frac{P^2}{4m})\frac{e^{-s(\beta-\tau)}}{(s-\varepsilon_1)(s-\varepsilon_2)}\label{eqprefdef}
\end{equation}
Notice that it is a function of the time difference $\tau"-\tau'$, as it should, but it is not obvious.
In order to get the Fourier transform, we deform the contour along the real axis. By doing so, we must take into account of the branch cut of $t_2(s)$ and the two poles at $s=\varepsilon_1$ and $\varepsilon_2$. For $a^{-1}\leq 0$ we find
\begin{multline}
{\mathcal P}\int_{\frac{P^2}{4 m}}^{+\infty}dx_1 \frac{e^{-(\beta-\tau)x_1}}{(x_1-\varepsilon_1)(x_1-\varepsilon_2)}\rho_2(x_1-\frac{P^2}{4 m})\\
+\frac{4\pi}{m^2 a(\varepsilon_1-\varepsilon_2)}\left(
\frac{e^{-(\beta-\tau)\varepsilon_1}}{\varepsilon_1-\frac{P^2}{4 m}+\frac{1}{m a^2}}-\left(\varepsilon_1\to \varepsilon_2  \right)
\right)
\end{multline}
where ${\mathcal P}$ is the Cauchy principal part. Notice that at unitarity, the second term vanishes.

Inside the interval $[\tau',\tau"]$, we obtain the two terms in the second parenthesis of Fig.\ref{figappsig2sum}. The first term, for given wave vectors ${\bf P}$, ${\bf k}_1$ and ${\bf k}_2$, is equal to
\begin{multline}
-T_2({\bf P}-{\bf k_1}+{\bf k},\tau)G^{(0,0)}({\bf k}_1,\tau)\delta({\bf k}_1-{\bf k}_2)(2\pi)^3=\\
e^{-\varepsilon_{k_1}\tau}T_2({\bf P}-{\bf k_1}+{\bf k},\tau)\delta({\bf k}_1-{\bf k}_2)(2\pi)^3\label{eqprefdef1}
\end{multline}
This can be written in term of inverse Laplace transform
\begin{multline}
\ds e^{-\tau s}T_2({\bf P}-{\bf k_1}+{\bf k},s-\varepsilon_{k_1})\delta({\bf k}_1-{\bf k}_2)(2\pi)^3\label{eqT2only}
\end{multline}
The second term is of the form (we do not write all wave vectors for simplicity)
\begin{multline}
\int dt_1dt_2 e^{-\varepsilon_{k_1}t_1}T_2(t_1)T_3(t_2)e^{-\varepsilon_{k_2}(\tau-t_1-t_2)}T_2(\tau-t_1-t_2)
\end{multline} 
which is again a convolution product and can be written as the inverse Laplace transform of the product of Laplace transforms
\begin{equation}
\ds e^{-s\tau}T_2(s-\varepsilon_{k_1}) T_3(s)T_2(s-\varepsilon_{k_2})\label{eqprefdef2}
\end{equation}
The next step is to deform the integration contours of the variable $s$ in Eqs.(\ref{eqT2only}) and (\ref{eqprefdef2}). This is done as before by introducing the spectral densities $\rho_2$ and $\rho_3$. 
The final result for the sum of the diagram (d)+(e)+(f) is
\begin{multline}
\Sigma^{(2,2,d+e+f)}(k,\tau)=-z^2 {\mathcal P}\int\frac{d{\bf P}_t d{\bf p'}_1 d{\bf p'}_2}{(2\pi)^9}\int_0^{+\infty}\hspace{-5.mm}dx'_1
\int_0^{+\infty}\hspace{-5.mm}dx'\\
e^{-\beta(x'_1+\frac{({\bf P}_t-{\bf k})^2}{4 m})}\frac{\rho_2(x'_1)}{(x'_1-\varepsilon'_1)(x'_1-\varepsilon'_2)}
e^{-\tau\left( x'+\frac{P_t^2}{6m}-x'_1-\frac{({\bf P}_t-{\bf k})^2}{4m}-\mu\right)}\\
\left[
\rho_2(x'-\frac{3}{4}\frac{(p'_1)^2}{m})\delta({\bf p}'_1-{\bf p}'_2)(2\pi)^3+\rho_3({\bf p}'_1,{\bf p}'_2;x')
\right]
\end{multline}
where we have made the changes of variables $x'=x-P_t^2/(6m)$, $x'_1=x_1-P^2/(4m)$ and $\varepsilon'_{1,2}=\varepsilon_{1,2}-P^2/(4m)$. If we take the Fourier transform and make the analytic continuation, we find Eqs.(\ref{eqsigdeffin1}), (\ref{eqsigdeffin2}).
\section{Proof of Eq.(\ref{eqT2T2})}\label{eqT2T2p}
\begin{figure}[h]
\begin{center}
\includegraphics[width=.65\linewidth]{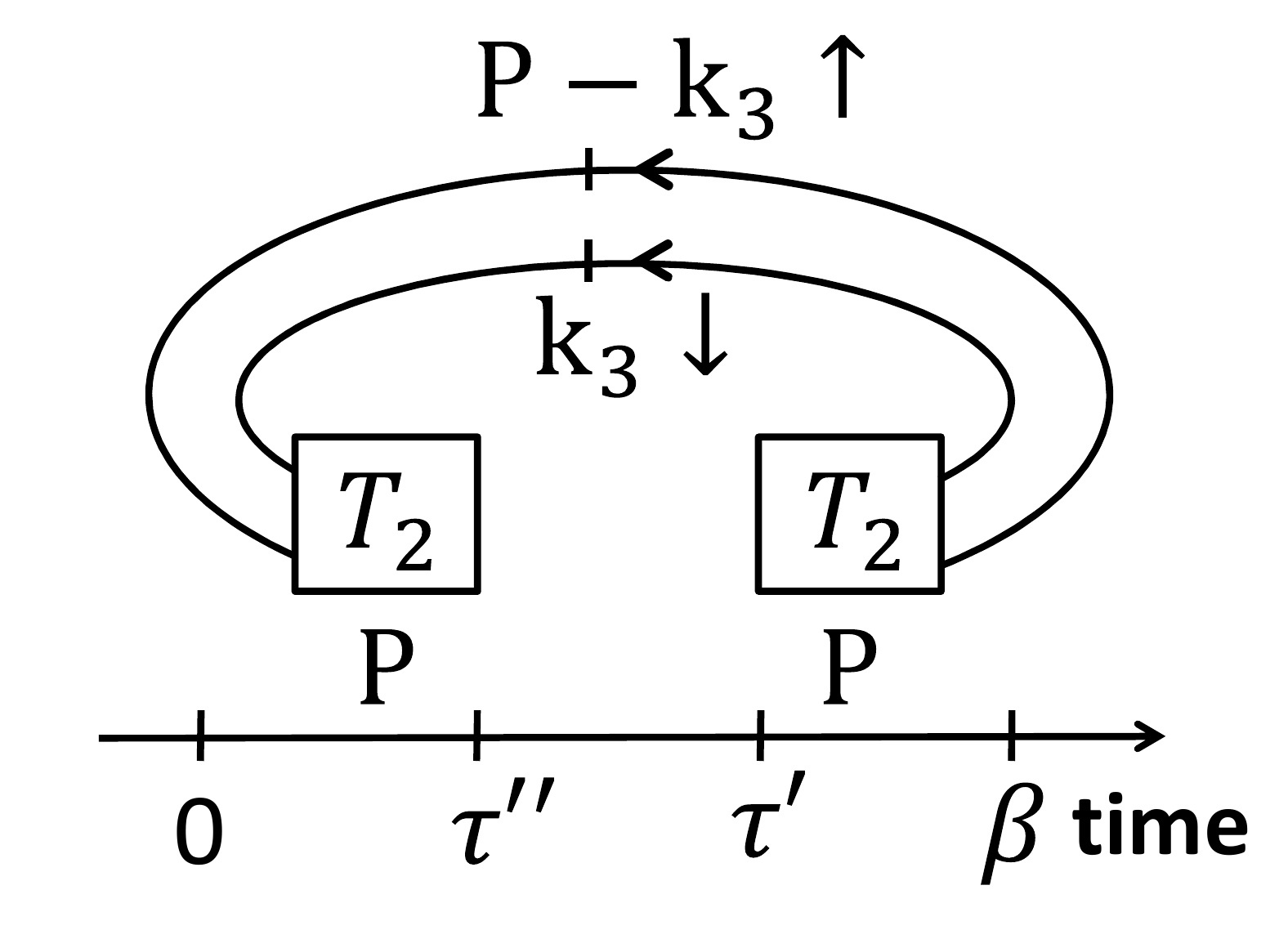}
\caption{ The diagram considered in Appendix \ref{appsigdef} for $\tau"<\tau'$.}
\label{diagT2T2}
\end{center}
\end{figure}

The diagram of Fig.\ref{diagT2T2} is denoted $I({\bf P};\tau",\tau')$, with $\tau"<\tau'$.
The two-particle vertex $\Gamma({\bf P},\tau)$ has an explicit imaginary time dependance $e^{2\mu\tau}$, due to the time evolution of the two created fermions. We define
\begin{equation}
\tilde{\Gamma}({\bf P},\tau)=e^{-2\mu\tau}\Gamma({\bf P},\tau)
\end{equation}
This new vertex can be expanded in terms of the fugacity $\tilde{\Gamma}({\bf P},\tau)=\sum_{n\geq 0}z^{n}\tilde{\Gamma}^{(n)}({\bf P},\tau)$, where $z=e^{\beta\mu}$.
Moreover we have the periodic boundary condition $\Gamma({\bf P},\tau<0)=\Gamma({\bf P},\beta+\tau)$, which yields  $\tilde{\Gamma}({\bf P},\tau<0)=z^2\tilde{\Gamma}({\bf P},\beta+\tau)$. Hence we have $\tilde{\Gamma}^{(n+2)}({\bf P},\tau<0)=\tilde{\Gamma}^{(n)}({\bf P},\beta+\tau)$. For $n=0$, it is easy to see that $\tilde{\Gamma}^{(0)}({\bf P},\tau)=T_2({\bf P},\tau)$, the $2$-particles $T$-matrix in vacuum. Moreover, it is also easy to show that for $\tau<0$, $-\tilde{\Gamma}^{(2)}({\bf P},\tau)$ is precisely given by the diagram of Fig.\ref{diagT2T2} (for $\tau<0$, one needs at least two slashed fermionic lines which go backward in time). Therefore we have
\begin{equation}
I({\bf P};\tau",\tau')=-T_2({\bf P},\beta+\tau"-\tau')
\end{equation}
\section{Structures in $\Im(\Sigma^{(2,2,a)}(k,\omega))$}\label{struct221}
In Eq.(\ref{eqsigmaF221}), we can make the change of variable ${\bf P}_t\to {\bf P}'$ with ${\bf P}_t=3{\bf P}'+3{\bf k}+3/2{\bf p}'_1$. In this way the imaginary part of $\Sigma^{(2,2,a)}(k,\omega)$ is given by
\begin{multline}
-\frac{1}{\pi}\Im(\Sigma^{(2,2,a)}(k,\omega))=3^3 z^2 \dpppp e^{-\beta\mathcal{E}_1}\\ 
\times\rho_{3}({\bf p'}_1,{\bf p'}_1;x)\delta(\tilde{\omega}-x+\frac{P'^2}{m}+\frac{3}{4}\frac{(p'_1)^2}{m})
\end{multline} 
with $\tilde{\omega}=\omega+\mu-k^2/(2m)$.
The integral on $x$ is performed. Due to the Dirac function, the integral on $x$ can be performed. We find $x=\tilde{\omega}+3/4(p'_1)^2/m+P'^2/m\equiv x_0\geq 0$. In this way, we find
\begin{multline}
-\frac{1}{\pi}\Im(\Sigma^{(2,2,a)}(k,\omega))=3^3 z^2\dpppbis e^{-\beta\mathcal{E}_1}\rho_{3}({\bf p'}_1,{\bf p'}_1;x_0)\\ \times\Theta(\tilde{\omega}+\frac{P'^2}{m}+\frac{3}{4}\frac{(p'_1)^2}{m})\label{eqsigma221ap}
\end{multline} 
In the unitary limit, $\rho_3({\bf p'}_1,{\bf p'}_1;x_0)$ is given by 
\begin{multline}
\rho_3({\bf p'}_1,{\bf p'}_1;x_0)=\frac{16\,\pi}{m^2}\{\mathcal{P}\frac{1}{\tilde{\omega}+\frac{(P')^2}{m}}\Im\left[t_3({\bf p'}_1, {\bf p'}_1;x_0) \right]\\
-\pi\delta(\tilde{\omega}+\frac{(P')^2}{m})\Re\left[t_3({\bf p'}_1, {\bf p'}_1;x_0) \right]\}\label{eqrho3ap}
\end{multline} The first term (respectively the second term) gives a contribution to $-1/\pi \Im(\Sigma^{(2,2,a)}(k,\omega))$ denoted $I_1(\tilde{\omega},k)$ (respectively  $I_2(\tilde{\omega},k)$). Hence we have
$-1/\pi \Im(\Sigma^{(2,2,a)}(k,\omega))=I_1(\tilde{\omega},k)+I_2(\tilde{\omega},k)$.

We first consider $I_2(\tilde{\omega},k)$. Due to the Dirac function $\delta(\tilde{\omega}+\frac{(P')^2}{m})$, we see that this term vanishes for $\tilde{\omega}>0$. Moreover, for $\tilde{\omega}<0$, the integration on the norm of ${\bf P}'$ gives a term, for small $|\tilde{\omega}|$, of order $\int_0 dP' (P')^2 \delta(-|\tilde{\omega}|+\frac{(P')^2}{m})\sim \sqrt{|\tilde{\omega}|}$. We come to the conclusion
\begin{equation}
I_2(\tilde{\omega},k)=\Theta(-\tilde{\omega})\sqrt{|\tilde{\omega}|}F(\tilde{\omega},k)
\end{equation}
where $F$ is a regular function of $\tilde{\omega}$ and $k$. This result is of course confirmed by our numerical calculations.

In the expression for $I_1(\tilde{\omega},k)$, $\tilde{\omega}$ enters in the imaginary part of $t_3$, in the $\Theta$ function in Eq.(\ref{eqsigma221ap}) and in the prefactor $1/(\tilde{\omega}+P'^2/m)$ in Eq.(\ref{eqrho3ap}). We first consider the $\tilde{\omega}$ dependence from this last term. We have found arguments which show that the other $\tilde{\omega}$ terms will be of order $\tilde{\omega}\ln(\tilde{\omega})$ and $\tilde{\omega}$, which are higher order terms. Following this method, we find for the difference
$I_1(\tilde{\omega},k)-I_1(0,k)$ (we take the mass $m=1$)
\begin{multline}
I_1(\tilde{\omega},k)-I_1(0,k)\approx 3^3 16\pi z^2\mathcal{P} \dpppbis e^{-\beta\mathcal{E}_1}\left[\frac{-\tilde{\omega}}{\tilde{\omega}+(P')^2}\right]\\
\times\frac{1}{(P')^2}
\Im\left[t_3({\bf p'}_1, {\bf p'}_1;3/4(p'_1)^2+P'^2+i\delta)\right]
\end{multline}
We see from this expression that for small $\tilde{\omega}$, the integration on $P'$ will be limited to values of order $\sqrt{|\tilde{\omega}|}$. At lowest order, we can set $P'=0$ in the expression in factor of $1/(\tilde{\omega}+(P')^2)$. The integration on $P'$ gives
$\mathcal{P}\int_0 dP' 1/(\tilde{\omega}+P'^2)=\pi/2\sqrt{\tilde{\omega}}\Theta(\tilde{\omega})$.
As a conclusion, we find 
\begin{equation}
I_1(\tilde{\omega},k)-I_1(0,k)=\sqrt{\tilde{\omega}}\Theta(\tilde{\omega})C+\cdots
\end{equation}
 The prefactor $C$ (which depends on $k$ and $\beta$) can be easily calculated once $t_3$ is known. The next order terms of order $\tilde{\omega}\ln(\tilde{\omega})$ and $\tilde{\omega}$ are fitted to reproduce the numerics with reasonable coefficients of order unity. We find that all the arguments of this Appendix give good analytical supports to our numerics.


\begin{thebibliography}{99}
\bibitem{erratumpra} There is a misprint in the sign of the factor $2^{-5/2}$ in Eq.(27) of Ref. \cite{pravirial}.
\end{thebibliography}
\end{document}